\documentstyle[12pt,epsfig,axodraw]{article}

\newcommand{\lsim}{\raisebox{-0.13cm}{~\shortstack{$<$ \\[-0.07cm] $\sim$}}~}
\newcommand{\gsim}{\raisebox{-0.13cm}{~\shortstack{$>$ \\[-0.07cm] $\sim$}}~}
\newcommand{\dx}{\mbox{\rm d}}
\newcommand{\ra}{\rightarrow}
\newcommand{\ee}{e^+e^-}
\newcommand{\tb}{\tan \beta}
\newcommand{\s}{\smallskip}
\newcommand{\nn}{\noindent}
\newcommand{\non}{\nonumber}
\newcommand{\beq}{\begin{eqnarray}}
\newcommand{\eeq}{\end{eqnarray}}

\newcommand{\ct}[1]{c_{\theta_#1}}
\newcommand{\st}[1]{s_{\theta_#1}}

\begin{document}

\vspace*{1cm} 

\begin{flushright}
PM/01--26\\
July 2001\\
\end{flushright}

\vspace*{0.9cm}

\begin{center}

{\large\sc {\bf Charged Higgs production from SUSY particle}}

\vspace*{0.4cm}

{\large\sc {\bf cascade  decays at the LHC}}

\vspace{0.7cm}

{\sc Aseshkrishna DATTA, Abdelhak DJOUADI,} 

\vspace{0.3cm}

{\sc  Monoranjan GUCHAIT} and {\sc Yann MAMBRINI}

\vspace{0.7cm}

$^1$ Laboratoire de Physique Math\'ematique et Th\'eorique, UMR5825--CNRS,\\
Universit\'e de Montpellier II, F--34095 Montpellier Cedex 5, France. 

\end{center} 

\vspace*{1cm} 

\begin{abstract}
\nn We analyze the cascade decays of the scalar quarks and gluinos of the
Minimal Supersymmetric extension of the Standard Model, which are abundantly
produced at the Large Hadron Collider, into heavier charginos and neutralinos
which then decay into the lighter ones and charged Higgs particles, and show
that they can have substantial branching fractions. The production rates of
these Higgs bosons can be much larger than those from the direct production
mechanisms, in particular for intermediate values of the parameter $\tan
\beta$, and could therefore allow for the detection of these particles. We also
discuss charged Higgs boson production from direct two--body top and bottom
squark decays as well as from two-- and three--body gluino decays.  
\end{abstract}

\newpage 

\subsection*{1. Introduction}

The most distinctive signature of an extended Higgs sector, compared to the
Standard Model (SM) where only one scalar doublet is needed to break the
electroweak symmetry leading to a single neutral Higgs particle, is the
discovery of charged Higgs bosons. For instance, in the Minimal Supersymmetric
(SUSY) extension of the Standard Model (MSSM) \cite{MSSM}, two Higgs doublets
are present, leading to the existence of a quintet of scalar particles: two
CP-even neutral Higgs bosons $h$ and $H$, a CP--odd neutral Higgs boson $A$ and
two charged Higgs particles $H^\pm$ \cite{HHG}. SUSY constraints on the Higgs
spectrum impose that the charged Higgs boson mass is related to the
pseudoscalar Higgs mass, $M^2_{H^\pm} = M_A^2+M_W^2$. With the present
experimental limit on the $A$ boson mass, $M_A \gsim 93.5$ GeV \cite{LEPh},
this leads to a bound $M_{H^\pm} \gsim 120$ GeV. In fact, in the popular
minimal Supergravity models (mSUGRA) with universal boundary conditions at 
the Grand Unification scale and where the electroweak symmetry breaking
is induced radiatively \cite{mSUGRA}, the $H^\pm$ boson, as well as the $H$ 
and $A$ neutral Higgs particles, tend to be rather heavy, with masses of the 
order of a few hundred GeV \cite{Tevatron}. The states are therefore 
kinematically accessible only at the LHC \cite{LHC}, at future $\ee$ linear 
colliders \cite{NLC} or muon colliders \cite{FMC}. \s

The discovery of $H^\pm$ bosons at the LHC through the standard processes is
rather difficult, if not impossible in some areas of the MSSM parameter space
\cite{Houches}.  This is mainly due to the fact that the production rates are
controlled by the charged Higgs boson Yukawa couplings to up-- and down--type
fermions. Using the notation of the first generation, the latter are given by
\cite{HHG}:
\beq
{g V_{ij} \over \sqrt{2} M_W} H^+ \left[\cot \beta \; m_{u} \; \bar u_i d_{jL}
+ \tan \beta \; m_d \; \bar u_i d_{jR} \right],
\label{coupling}
\eeq
where $V_{ij}$ is the CKM matrix with $u_i=u,c,t$, $d_i=d,s,b$ and $\tb =v_2/
v_1$ is the ratio of the vacuum expectation values of the two Higgs doublets 
needed to break the electroweak symmetry in the MSSM.  
For values $\tan \beta >1$, as is the case in the MSSM, the couplings to
down--type (up--type) fermions are enhanced (suppressed). Only the couplings to
the top and bottom isodoublet quarks are therefore important, in particular for
small $\sim 1$ and large $\sim m_t/m_b$ values\footnote{Interestingly,
these two regions of $\tan\beta$ are favored by Yukawa coupling unification. 
However, the experimental bound on the lightest $h$ boson at LEP2, $M_h \gsim
113.5$ GeV \cite{LEPh} in the decoupling regime where the $H^\pm$ bosons 
are heavy, rules out the low $\tb$ scenario.}. \s

A light charged Higgs particle can be searched for at the Tevatron in top
decays through the process $p\bar{p}\rightarrow t \bar{t}$, with at least one
of the top quarks decaying via $t\rightarrow H^+ b$, leading to a surplus of
$\tau$ leptons  due to the $H^\pm \to \tau^\pm \nu$ decay, an apparent
breakdown of $\tau$--$\mu$--$e$ universality. For small and large values of
$\tb$, the branching ratio BR($t\rightarrow H^+ b$) is large and would allow
for the detection of the signal at the Tevatron \cite{Tev-rep}. The situation
for intermediate values of $\tb$, where the $H^- t\bar{b}$ coupling is small,
leading to a tiny $t\rightarrow H^+ b$ branching ratio, is rather difficult and
should await for the LHC. Indeed, detailed analyses of the ATLAS and CMS
collaborations have shown that the entire range of $\tan \beta$ values should
be covered for $M_{H^\pm} \lsim m_t$ using this process \cite{LHC}.  \s

For charged Higgs bosons with masses $M_{H^\pm} >m_t$, the two production 
mechanisms which potentially have sizeable cross sections at the LHC are 
\cite{prod1,prod2}:
\beq
pp \to & gb (g\bar{b}) \to tH^- \; (\bar{t}H^+) & 
\nonumber \\
pp \to & gg/q\bar{q} \to tH^- \bar{b} + \bar{t}H^+ b & 
\eeq
The signal cross section from the $2 \to 2$ mechanism $gb \to tH^-$, where the
$b$ quark is obtained from the proton, is 2--3 times larger than the $2 \to 3$
process $gg/q\bar{q} \to t\bar{b} H^-$, where the $H^-$ boson is radiated from
a heavy quark line. When the decays $H^+ \to t\bar{b}$ and $t \to Wb$ take
place, the first process gives rise to 3 $b$--quarks in the final state while
the second one gives 4 $b$--quarks; both processes contribute to the inclusive
production where at most 3 final $b$--quarks are tagged\footnote{However, the
two processes have to be properly combined to avoid double counting of the
contribution where a gluon gives rise to a $b\bar{b}$ pair that is collinear to
the initial proton. The cross section of the inclusive process in this case is
mid--way between those of the two production mechanisms eqs.~(2)
\cite{Houches}.}.  However, the cross sections are rather small: even for the
extreme values $\tb =2$ and $40$, they hardly reach the level of a picobarn for
a charged Higgs boson mass $M_{H^\pm}=200$ GeV. For intermediate values of
$\tb$ and/or larger $H^\pm$ masses, the cross sections are too small for these
processes to be useful. For instance, for the value $\tb=10$, the cross section
is below the level of a few femtobarn for $M_{H^\pm} \gsim 250$ GeV.\s

Other mechanisms for $H^\pm$ production at hadron colliders are the Drell--Yan
type process for pair production through $\gamma$ and $Z$ boson exchange, $q
\bar{q} \to H^+H^-$ \cite{qqH+}, the gluon--gluon fusion process for pair
production, $gg \to H^+H^-$ \cite{ggHH+}, and the associated production process
with $W$ bosons in $gg$ fusion and $q\bar{q}$ annihilation, $q\bar{q}, gg \to
H^\pm W^\mp $ \cite{ppWH+}. However, the production rates are rather small at
the LHC: for the quark--antiquark processes because of the low quark
luminosities at high energies and for the gluon--gluon fusion processes because
they are induced by loops of heavy quarks (and squarks) and are thus suppressed
by additional coupling factors. The cross sections are at the femtobarn level
for large enough charged Higgs boson masses, and are therefore too low to be
easily useful in the complicated and hostile environment of hadron colliders.\s

In this paper, we show that there is a potentially large source of the $H^\pm$
bosons of the MSSM at the LHC: the cascade decays of squarks and gluinos, which
are abundantly produced in $pp$ collisions, thanks to their strong
interactions\footnote{These decays have been discussed in the past \cite{Baer}
for charged Higgs bosons with masses below $\sim 150$ GeV and which can thus
also be produced in top quark decays.}. Squarks and gluinos can decay into
the heavy charginos and neutralinos, $\chi_2^\pm, \chi_3^0$ and $\chi_4^0$, and
if enough phase space is available, the latter could decay into the lighter
charginos/neutralinos, $\chi_1^\pm, \chi_1^0$ and $\chi_2^0$, and charged Higgs
bosons\footnote{We will not specifically consider here the similar production
of neutral Higgs particles, $\Phi=h,H,A$, since the later can be produced
abundantly in standard processes such as as gluon--gluon fusion, $gg \to \Phi$
and associated production with heavy quarks $gg, q\bar{q} \to b\bar{b}\Phi,
t\bar{t}\Phi$ \cite{LHC}.}:
\begin{eqnarray} 
pp \to \tilde{g} \tilde{g} , \tilde{q} \tilde{q}, \tilde{q} \tilde{g}  & \to &
\chi_2^\pm, \chi_3^0, \chi_4^0 + X \non \\
& \to &  \chi_1^\pm, \chi_2^0, \chi_1^0 + H^\pm \ +X 
\eeq 
These processes are similar to the ones with cascade decays of strongly
interacting SUSY particles into the next--to--lightest neutralino $\chi_2^0$
which then decays into the lightest $h$ boson and the lightest neutralino
[which is expected to be the lightest SUSY particle in the MSSM], a process
which has been discussed in the literature; see for instance Ref.~\cite{LHC2}.  

Charged Higgs bosons could also be searched for, if kinematically possible, in 
the direct decays of heavy third generation squarks into their lighter 
partners, 
\beq
\tilde{Q} \to \tilde{Q}' H^\pm  \ \ {\rm with} \ \  \tilde{Q}, \tilde{Q}' =
\tilde{t},\tilde{b}
\eeq
or in direct gluino three--body decays into heavy quarks, their partners 
squarks and $H^\pm$ bosons, 
\beq
\tilde{g} \to Q' \tilde{Q} H^\pm \ \ {\rm with} \ \  \tilde{Q} =
\tilde{t},\tilde{b} 
\eeq

The remainder of the paper is organised as follows. In the next section, we
discuss the production cross sections of squarks and gluinos at the LHC as well
as their main decay modes. In section 3, we analyse the decays of the heavy 
charginos and neutralinos into the lighter charginos/neutralinos and Higgs or 
gauge bosons and estimate the production cross sections for $H^\pm$ bosons at 
the LHC in some specific scenarii. In section 4, we discuss the direct 
production of $H^\pm$ bosons in the two--body decays of top and bottom squarks 
and three--body decays of gluinos. A short conclusion is given in section 5.

\subsection*{2. Production and decay modes of Squarks and Gluinos}

\subsection*{2.1 Production cross sections} 

In proton--proton collisions, gluino pairs are produced through $q\bar{q}$
annihilation and gluon--gluon fusion, $q\bar{q}, gg \to \tilde{g} \tilde{g}$. 
Squark pairs can be produced through $t$--channel exchange, $qq  \to \tilde{q}
\tilde{q}$, while squark--antisquark pairs are produced in both $q\bar{q}$
annihilation and gluon--gluon fusion. Finally, mixed squark--gluino production
proceeds through $s$-- and $t$-- channel $gq$ annihilation. The production
cross sections at the tree--level are given in Ref.~\cite{susyprod}. They have
however, to be supplemented by the next--to--leading order QCD radiative
corrections, which stabilize the theoretical predictions. The $K$--factors can
be rather large, ranging from $K= \sigma_{\rm NLO}/\sigma_{\rm LO} \sim 1$ to 2
\cite{Kfactor}, depending on the ratio of the squark to gluino masses. Taking
for the renormalisation and factorization scale the average mass $m$ of the two
produced sparticles results in a conservative estimate of the total production
cross section. We will therefore not include the $K$--factors in our analysis
and set the scale at which the cross section is evaluated to be the average
mass of the final particles. \s

The cross sections for pair production and associated production of squarks and
gluinos are shown in Fig.~1 as functions of the squark and gluino masses.  The
CTEQ3L \cite{CTEQ} parameterization of the parton densities has been used.  In
Fig.~1a, we choose $m_{\tilde{g}}= m_{\tilde{q}}$ while in Figs.~1b and 1c, we
take respectively, $m_{\tilde{q}} = 1.2 m_{\tilde{g}}$ and $m_{\tilde{g}}= 1.2
m_{\tilde{q}}$. We assume the left-- and right--handed scalar partners of the
light quarks, including the $b$--quark, to be degenerate in mass and we sum
the individual cross sections. We show separately in Fig.~1a, the total cross
section for the pair production of the lightest top squark, $\sigma_{q\bar{q}},
gg \to \tilde{t}_1 \tilde{t}_1^*)$, which depends only on $m_{\tilde{t}_1}$
at the tree--level. \smallskip

\begin{figure}[htbp]
\begin{center}
\vskip-4cm
\hskip-1cm\centerline{\epsfig{file=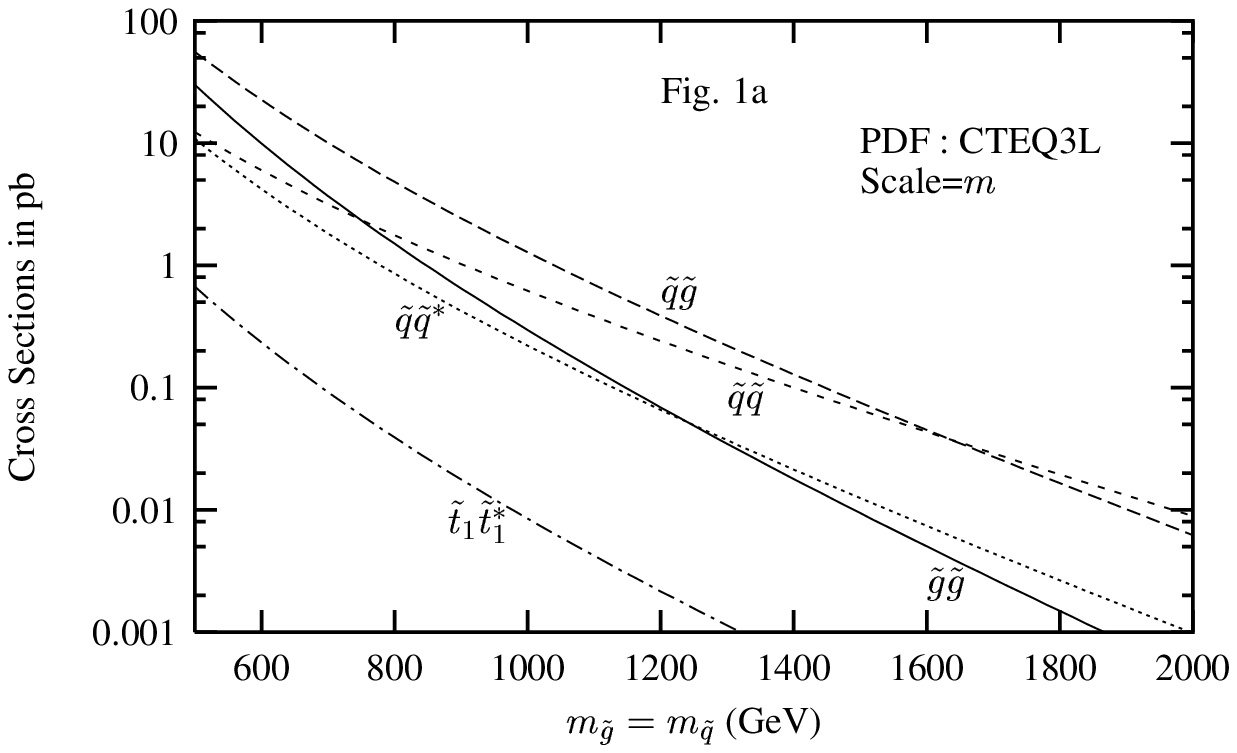,width=17cm}}
\vskip-17.5cm
\hskip-1cm\centerline{\epsfig{file=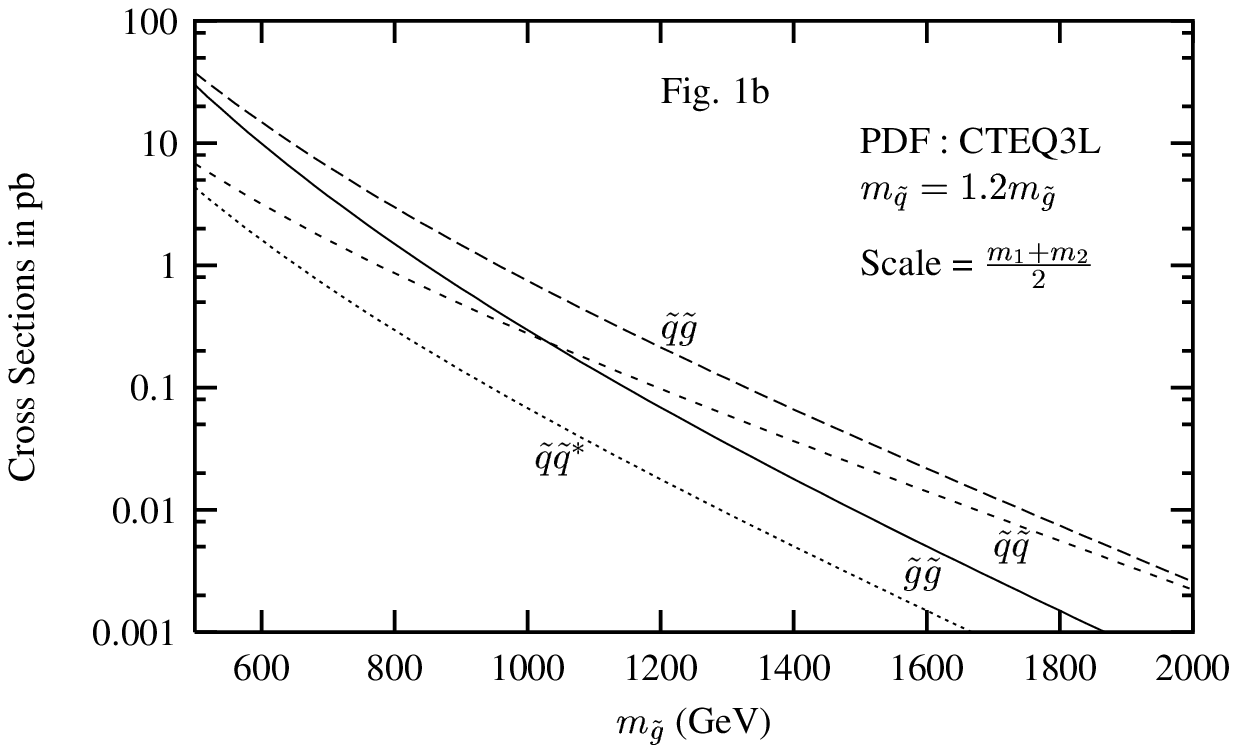,width=17cm}}
\vskip-17.5cm
\hskip-1cm\centerline{\epsfig{file=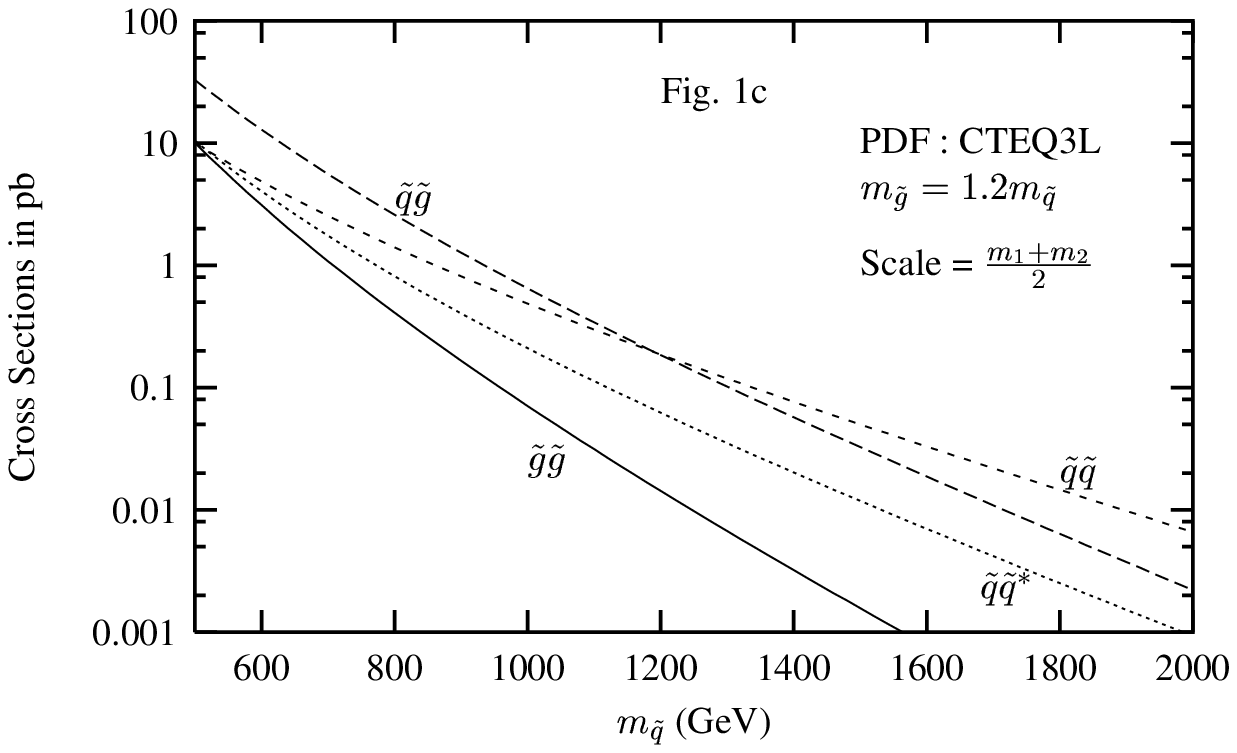,width=17cm}}
\vskip-14cm
\caption{Pair and associated production cross sections at the lowest order 
for the strongly interacting SUSY particles at LHC with $\sqrt{s}=14$ TeV
as a function of the final state masses for representative choices of squark
and gluino masses. $\sigma_{\tilde{q} \tilde{q}}$ and $\sigma_{\tilde{q} 
\tilde{g}}$ include the contributions from $\sigma_{\tilde{q} 
\tilde{q}^*}$ and $\sigma_{\tilde{q} \tilde{g}}$ respectively;  
$\sigma_{\tilde{t}_1 \tilde{t}_1^*} $ is presented only in Fig. 1a.}
\end{center}
\end{figure}

We see that the largest cross section is due to squark and gluino associated
production, in which squarks and anti--squarks of the five light flavors have
been added. It ranges from $\sigma_{\tilde{\bar{q}} \tilde{g}} \sim 50$ pb for
a gluino mass $m_{\tilde{g}} \sim 500$ GeV to ${\cal O}$ (1 pb) for
$m_{\tilde{g}} \sim 1$ TeV. In the low mass range, it is followed by the cross
section for gluino (squark) pair production if the gluino is lighter (heavier)
than squarks.  The total cross sections decrease quickly with increasing final
state  masses. In the case of top squarks, the total production cross section
is an order of magnitude smaller than the one for squark--antisquark pairs,
$\sigma_{\tilde{t}_1 \tilde{t}_1^*} \sim \frac{1}{12} \sigma_{\tilde{q} 
\tilde{q}^*}$ if all squarks, including $\tilde{t}_1$, have the same mass; the cross
section for the production of $\tilde{t}_2$, which is expected to be heavier
than $\tilde{t}_1$ because of the generally large mixing in the stop sector, is
smaller than $\sigma_{\tilde{t}_1 \tilde{t}_1^*}$. \s

Summing up all cross sections, one obtains $\sigma_{\tilde{q}+\tilde{g}} \sim
110$ (3) pb for $m_{\tilde{g}}\sim m_{\tilde{q}}\sim 500$ (1000) GeV. This means
that with the expected integrated luminosity at the LHC, $\int {\cal L} \sim
300$ fb$^{-1}$, a total of $3 \cdot 10^{7}$ to  $10^{5}$ events can be
collected by the ATLAS and CMS collaborations in a course of few years.  It is
therefore tempting to use this very large sample of events to look for
particularly interesting decay channels of squarks and gluinos which, even if
they have branching ratios below the percent level, would still lead to a
rather large number of final state events that could be studied in detail.  

\subsection*{2.2 Squark Decays} 

The main decay modes of squarks, if they are lighter than the gluino, will be
into their partner quarks and neutralinos, $\tilde{q}_i \ra q \chi^0_j$
[$j$=1--4], as well as quarks and charginos, $\tilde{q}_i \ra q' \chi^\pm_j$
[$j$=1--2]. Taking into account the mass of the final quark, which would be
appropriate in the case of top squark decays, the partial decay widths are
given at the tree--level by [the QCD corrections to these decay modes have been
calculated in Ref.~\cite{C1}]:
\begin{eqnarray}
\Gamma( \tilde{q}_i \ra q \chi_j^0) &=& \frac{\alpha \lambda^{\frac{1}{2}} 
(\mu_{q}^2,\mu_{\chi_j^0}^2)}{4}m_{\tilde{q}_i}
             \bigg[ ( {a^{\tilde q}_{ij}}^2 + {b^{\tilde q}_{ij}}^2 ) 
                     ( 1 - \mu_{q}^2 - \mu_{\chi_j^0}^2 )
                    - 4 a^{\tilde q}_{ij} b^{\tilde q}_{ij} \mu_{q} 
             \mu_{\chi_j^0} \epsilon_{\chi_j} \bigg] \nonumber \\
\Gamma(\tilde{q}_i \ra q' \chi_j^\pm) &=& \frac{\alpha \lambda^{\frac{1}{2}}
(\mu_{q'}^2,\mu_{\chi_j^+}^2)}{4} m_{\tilde{q}_i}
             \bigg[ ( {a^{\tilde q}_{ij}}^2 + {b^{\tilde q}_{ij}}^2 )  
                     (1 - \mu_{q'}^2 - \mu_{\chi_j^+}^2 )
                    - 4 a^{\tilde q}_{ij} b^{\tilde q}_{ij} \mu_{q'} 
\mu_{\chi_j^+}              \bigg] \
\end{eqnarray}
where $\lambda(x,y)=1+x^2+y^2-2(xy+x+y)$ is the usual two--body phase space
function with the reduced masses $\mu_X=m_X/m_{\tilde{q}_i}$ and
$\epsilon_{\chi_j}$ is the sign of the eigenvalue of the neutralino $\chi_j^0$.
In terms of $s_W^2=1-c_W^2\equiv \sin^2\theta_W$, the squark electric charge,
weak isospin and the mixing angle $\theta_q$ which turns the left-- and
right--handed states into the mass eigenstates $\tilde{q}_1 = c_{\theta_q}
\tilde{q}_L + s_{\theta_q} \tilde{q}_R$ and  $\tilde{q}_2 = -s_{\theta_q}
\tilde{q}_L + c_{\theta_q} \tilde{q}_R$, one has for the couplings among
neutralinos, quarks and squarks: 
\begin{eqnarray}
\left\{ \begin{array}{c} a^{\tilde q}_{j1} \\  a^{\tilde q}_{j2} \end{array} 
\right\}
        &=&   -\frac{m_q r_q}{\sqrt{2} M_W s_W}
\left\{ \begin{array}{c} \st{q} \\ \ct{q} \end{array} 
\right\}
         - e^q_{Lj}\, \left\{ \begin{array}{c} \ct{q} \\ -\st{q} \end{array} 
\right\} \nonumber \\
\left\{ \begin{array}{c} b^{\tilde q}_{j1} \\  b^{\tilde q}_{j2} \end{array} 
\right\}
        &=&  -\frac{m_q r_q}{\sqrt{2} M_W s_W }
\left\{ \begin{array}{c} \ct{q} \\ -\st{q} \end{array} 
\right\}
         - e^q_{Rj}\, \left\{ \begin{array}{c} \st{q} \\ \ct{q} \end{array} 
\right\}
\end{eqnarray}
with $r_u= Z_{j4}/ \sin \beta$ and $r_d=Z_{j3}/\cos \beta$ for up and 
down--type fermions, and  
\beq
e^q_{Lj}  =  \sqrt{2}\left[ e_q \; Z_{j1}'
              + \left(I_q^3 - e_q \, s_W^2 \right)
                 \frac{1}{c_W s_W}\;Z_{j2}' \right]  \ , \  
e^q_{Rj}  = -\sqrt{2} \, e_q \left[ Z_{j1}'
              - \frac{s_W}{c_W} \; Z_{j2}' \right]  
\eeq
while for the couplings among charginos, fermions and sfermions, 
$\tilde q_i-q'-\chi_j^+$, one has for up--type and down--type sfermions: 
\beq
\left\{ \begin{array}{c} a^{\tilde u}_{j1} \\ a^{\tilde u}_{j2} \end{array} 
\right\} & = &
   \frac{V_{j1}}{s_W} \,
\left\{ \begin{array}{c} -\ct{u} \\ \st{u} \end{array} \right\}
 + \frac{m_u\,V_{j2}}{\sqrt{2}\,M_W s_W\,s_\beta}
\left\{ \begin{array}{c} \st{u} \\ \ct{u} \end{array} \right\} 
\non \\
\left\{ \begin{array}{c} b^{\tilde u}_{j1} \\ b^{\tilde u}_{j2} \end{array} 
\right\} & = &
    \frac{m_d\,U_{j2}}{\sqrt{2}\,M_W s_W \, c_\beta}
\left\{ \begin{array}{c} \ct{u} \\ -\st{u} \end{array} \right\} \\
\left\{ \begin{array}{c} a^{\tilde d}_{j1} \\ a^{\tilde d}_{j2}
\end{array} \right\} & = &
   \frac{U_{j1}}{s_W} \,
\left\{ \begin{array}{c} -\ct{d} \\ \st{d} \end{array} \right\}
 + \frac{m_d\,U_{j2}}{\sqrt{2}\,M_W s_W \,c_\beta}
\left\{ \begin{array}{c} \st{d} \\ \ct{d} \end{array} \right\} 
\non \\
\left\{ \begin{array}{c} b^{\tilde d}_{j1} \\ b^{\tilde d}_{j2} \end{array} 
\right\} & = &
   \frac{m_u\,V_{j2}}{\sqrt{2} \,M_W s_W \,s_\beta}
\left\{ \begin{array}{c} \ct{d} \\ -\st{d} \end{array} \right\}
\eeq
In these expressions, $U,V$ and $Z$ are the (real) diagonalizing matrices for 
the chargino and neutralino states \cite{C0} with: 
\begin{eqnarray}
Z'_{i1}=Z_{i1}c_W +Z_{i2} s_W \ , \ 
Z'_{i2}=-Z_{i1} s_W + Z_{i2} c_W \ , \ 
Z'_{i3}=Z_{i3} \  , \ 
Z'_{i4}=Z_{i4} 
\end{eqnarray} 
If squarks are heavier than the gluino, they can also decay into gluino--quark
final states, for which the partial decay width is given by: 
\beq
\Gamma( \tilde{q}_i \ra q \tilde{g}) &=& \frac{2 \alpha_s \lambda^{\frac{1}{2}} 
(\mu_{q}^2,\mu_{\tilde{g}}^2)}{3} m_{\tilde{q}_i} 
             \bigg[ 1- \mu_{q}^2 - \mu_{\tilde{g}}^2 - 4
             a_{i\tilde{g}}^{\tilde{q}} b_{i\tilde{g}}^{\tilde{q}} 
               \mu_{q} \mu_{\tilde{g}} \bigg] 
\eeq
with the same notation as previously and the squark--quark--gluino coupling are:
\beq
a_{1\tilde{g}}^q = b_{2\tilde{g}}^q = \sin \theta_q \ \ , \ \ 
a_{2\tilde{g}}^q = - b_{1\tilde{g}}^q = \cos \theta_q 
\eeq
We will now discuss some scenarii for these decay modes, starting with the 
case of the scalar partners of light quarks and continuing with the special 
case of top squarks. 

\subsubsection*{2.2.1 The case of the scalar partners of light quarks}

If squarks are heavier than the gluino, they will decay most of the time into
quark plus gluino final states.  This is essentially due to the fact that these
are strong interaction decays, compared to their weak interaction decays into
charginos and neutralinos. The large value of the strong coupling constant,
$\alpha_s/\alpha \sim 10$, makes the decays into gluinos an order of magnitude
larger. In the opposite case, $m_{\tilde{q}} < m_{\tilde{g}}$, the
right--handed squarks will decay [for small quark masses] only into quarks and
neutralinos, while left--handed squarks decay into both charginos and
neutralinos. Two scenarii are possible, depending on the chargino/neutralino
textures: \s

(i) Gaugino--limit: If the lighter chargino and neutralinos are gaugino--like,
that is if the higgsino mass parameter is much larger than the wino and bino
mass parameters, $|\mu |\gg M_{1,2}$ with $M_{1}$ and $M_{2}$ related by the
GUT constraint $M_2 \sim 2 M_1$, the masses are such that $m_{\chi_2^0} \sim
m_{\chi_1^\pm} \sim 2 m_{\chi_1^0} \sim M_2$ while the heavy chargino and
neutralinos have masses, $m_{\chi_3^0} \sim m_{\chi_4^0} \sim m_{\chi_2^\pm}
\sim |\mu|$. Squarks will have then the tendency to decay into the lighter
ino states not only because of a more favorable phase space, but also because
for the partners of the light quarks, the higgsino component of the
quark--squark--ino coupling [which is proportional to $m_q$, see the previous
formulae for the couplings] is very small.  Therefore squarks will decay
dominantly into the lighter charginos and neutralinos. \s

(ii) Higgsino--limit: If, the lighter chargino and neutralinos are
higgsino--like, that is $M_{1,2} \gg |\mu|$, the trend is reversed and one
would have the mass hierarchies, $m_{\chi_2^0} \sim m_{\chi_1^\pm} \sim
m_{\chi_1^0} \sim |\mu|$ and $2m_{\chi_3^0} \sim m_{\chi_2^\pm} \sim
m_{\chi_4^0} \sim M_2$.  If allowed by phase space, i.e. for $m_{\tilde{q}} >
M_2$, (left--handed) squarks will decay into all possible neutralino and
chargino combinations in principle.  However, because the higgsinos couple
proportionally to the quark masses, the partial decay widths into the lighter
chargino and neutralinos are tiny, and squarks will dominantly decay into the
heavier chargino and neutralinos, $\tilde{q} \to q' \chi_2^\pm, q\chi_3^0$ and
$q \chi_4^0$. [In the mixed region, $M_2 \sim |\mu|$ decays into all charginos
and neutralinos are possible if phase space allowed.] \s

The two scenarii (i) and (ii) are exemplified in the upper and lower panels of
Fig.~2, respectively, where we display the ``average" squark branching ratios
into gluinos, charginos and neutralinos as a function of the squark mass for
the value $\tb=10$. This ``average" branching ratio is defined as the branching
fraction of the decay of any squark [except for top squarks] into a given
neutralino or chargino [or gluino] final state; this means that we sum all
possibilities for left-- and right--handed squarks as well as for up--type and
down--type squarks keeping proper track of flavors and chirality to reach a
given final state. [Note that here we neglect the small effect of the
$b$--quark Yukawa coupling which is not enough enhanced for the value of $\tb$
we are using, and treat the $\tilde{b}$ squark in the same footing as the first
and second generation squarks].  The gluino mass is fixed to $m_{\tilde{g}}=3
M_2$, which is typically the case in models with unified gaugino masses at the
GUT scale which lead to the tree--level relation $m_{\tilde{g}} \sim M_3 \sim 3
M_2$ at the low--energy scale. \smallskip

\begin{figure}[htbp]
\begin{center}
\vskip-5cm
\hskip-1cm\centerline{\epsfig{file=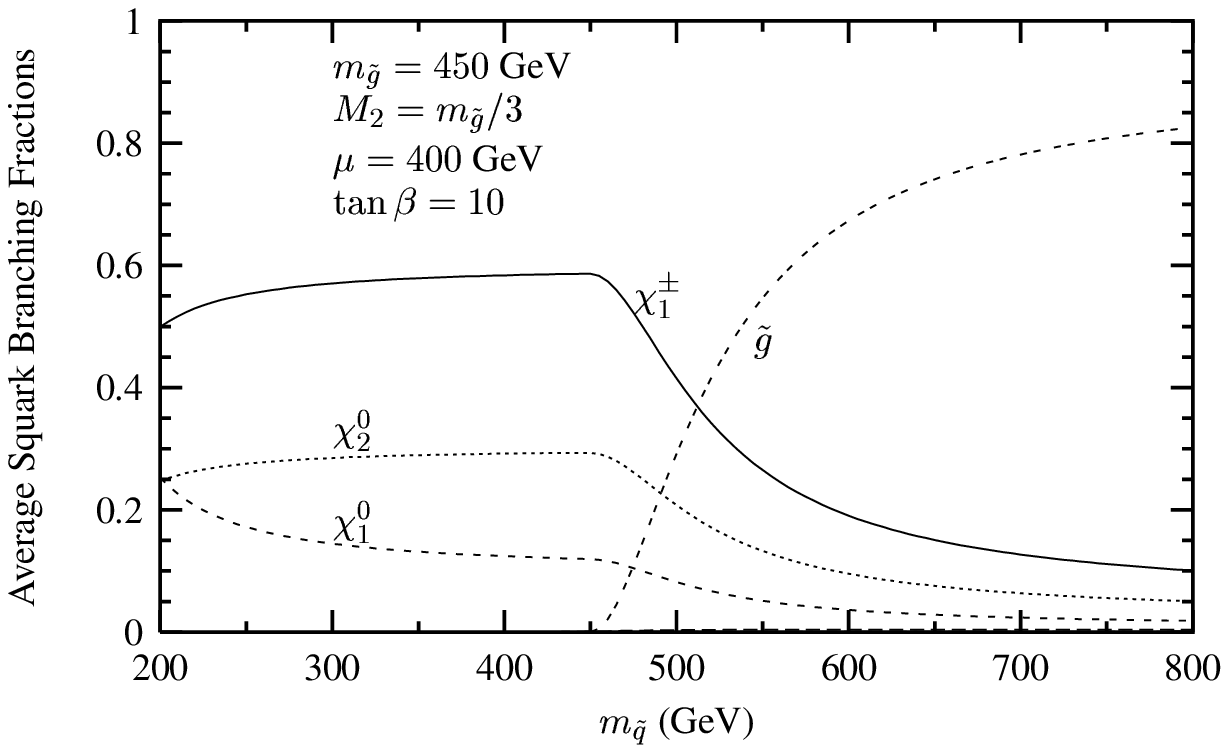,width=23cm}}
\vskip-24cm
\hskip-1cm\centerline{\epsfig{file=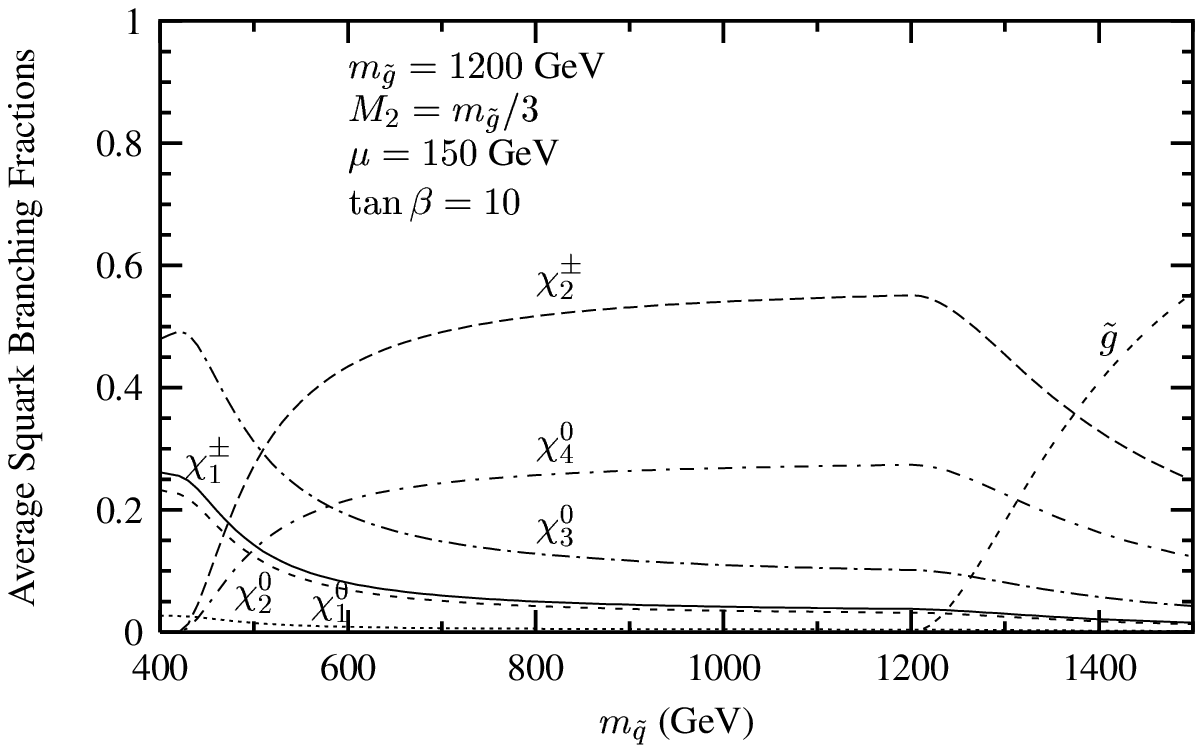,width=23cm}}
\vskip-19cm
\end{center}
\caption{Branching ratios of squarks decaying into gluinos, charginos and 
neutralinos as a function of their mass for $\tb=10$ and  $m_{\tilde{g}}=450$ 
GeV (top) 1.2 TeV (bottom). The gaugino mass is fixed to $M_2= m_{\tilde{g}}/3$
while the value of the parameter $\mu$ is 400 GeV and 150 GeV in scenarii (top)
and (bottom) respectively.}
\end{figure}

We see that in scenario (i), where the wino and higgsino mass parameters are
set to $M_2=150$ GeV and $\mu=400$ GeV, squarks will decay mainly into the
lighter chargino $\chi_1^\pm$ and neutralinos $\chi_{1,2}^0$ in the low mass
range, i.e. $m_{\tilde{q}} \lsim 450$ GeV, with branching ratios of the order
of 50\% for the charged and neutral decay channels, while the strong
interaction $\tilde{q} \to q \tilde{g}$ decay channel becomes by far dominating
above the gluino mass threshold, with a branching ratio reaching 90\% at large
squark masses. In scenario (ii), where the higgsino and wino mass parameters
are set to $\mu=150$ GeV and $M_2=400$ GeV, the gluino is much heavier than the
squarks, $m_{\tilde{g}} \gsim 1.2$ TeV, and squarks will decay almost 100\% of
the time into the heavier chargino and neutralinos, with a dominance at large
squark masses, of the charged decay mode, $\tilde{q} \to q' \chi_2^\pm$, which
reaches a branching fraction of 50\% due to the usual dominance of the charged
currents over the neutral currents. \s

Thus, there exist situations in which heavy squarks can
decay into the heavier chargino and neutralinos with significant rates. 

\subsubsection*{2.2.2 The case of the top squarks}

The case of top squarks is special: because of the large value of top quark
Yukawa coupling, there is a sizeable splitting between the two states
$\tilde{t}_2$ and $\tilde{t}_1$, and the latter is in general much lighter than
all other squarks. In some cases, gluinos can be lighter than the scalar
partners of light squarks, but heavier than $\tilde{t}_1$. In particular, one
can have the mass hierarchy $m_{\tilde{q}} \ge m_{\tilde{g}} \ge
m_{\tilde{t}_1}+m_t$ and the squarks $\tilde{q}$ will decay almost exclusively
into gluinos and quarks as discussed previously, and the gluinos will decay
into the only two--body decay mode which is allowed, i.e. $\tilde{g} \to t
\tilde{t}_1$. This means that all strongly interacting particles produced at
the LHC could decay dominantly into top and stop final states. \s

The branching ratios for $\tilde{t}_1$ decays into $\chi^+ b$ and $\chi^0t$
final states are shown in Fig.~3 for a mass of $m_{\tilde{t}_1} =600$ GeV and
$\tb=10$. In the upper (lower) panel, they are shown as functions of $\mu
(M_2)$ for a fixed value of $M_2=150$ GeV ($\mu=400$ GeV). As can be seen,
there are large regions where the decay $\tilde{t}_1 \to b\chi_2^+$ is
dominating. In particular, one can see that for small values of $\mu$ $(\sim
150$ GeV) and large values of $M_2$ ($\sim 400$ GeV), i.e. in the higgsino
region, BR($\tilde{t}_1 \to b\chi_2^+)$ can reach the level of 50\%.  But in
contrast to the scalar partners of light quarks, the branching ratio can also
be large in the gaugino region, $M_2=150$ GeV and $\mu=400$ GeV, since now the
couplings of top/stop states to higgsinos are enhanced by $m_t$. The decays
into neutralinos, $\tilde{t}_1 \to t \chi_{3,4}^0$ are also sizeable in the
gaugino region, but the phase space is limited because of the large value of
$m_t$.  

\begin{figure}[htbp]
\begin{center}
\vskip-5cm
\hskip-1cm\centerline{\epsfig{file=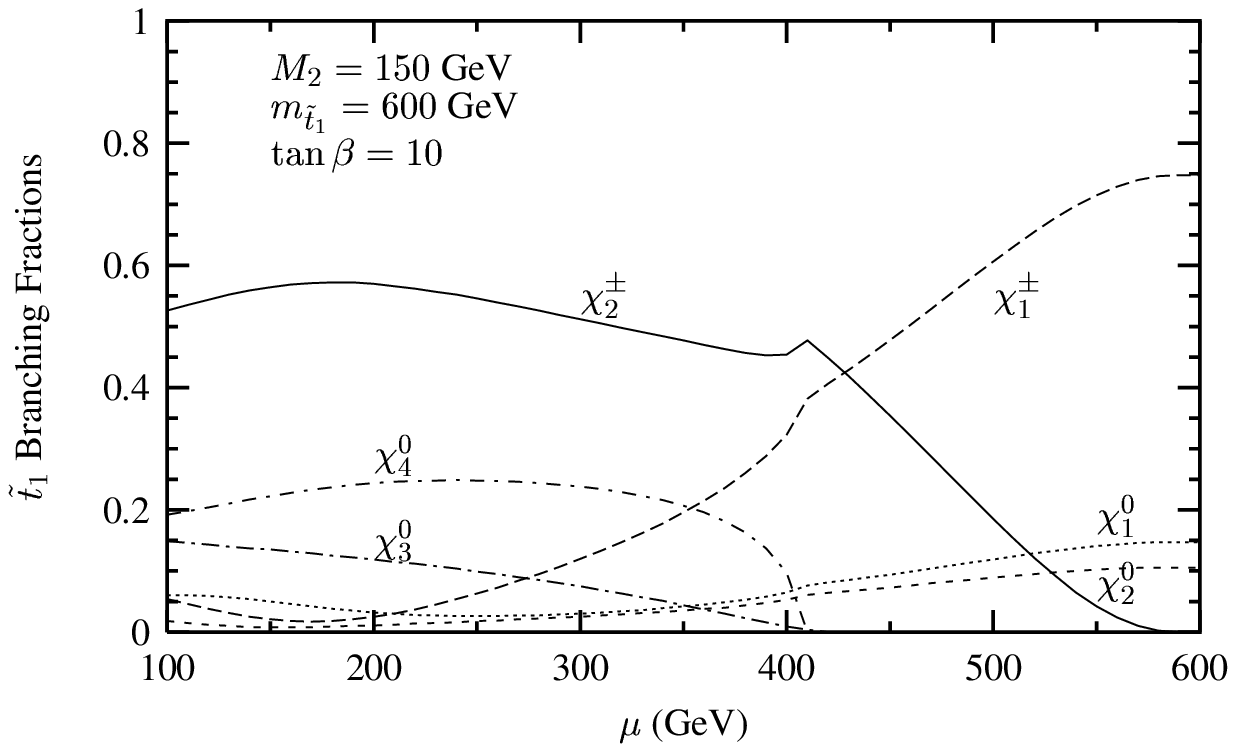,width=23cm}}
\vskip-24cm
\hskip-1cm\centerline{\epsfig{file=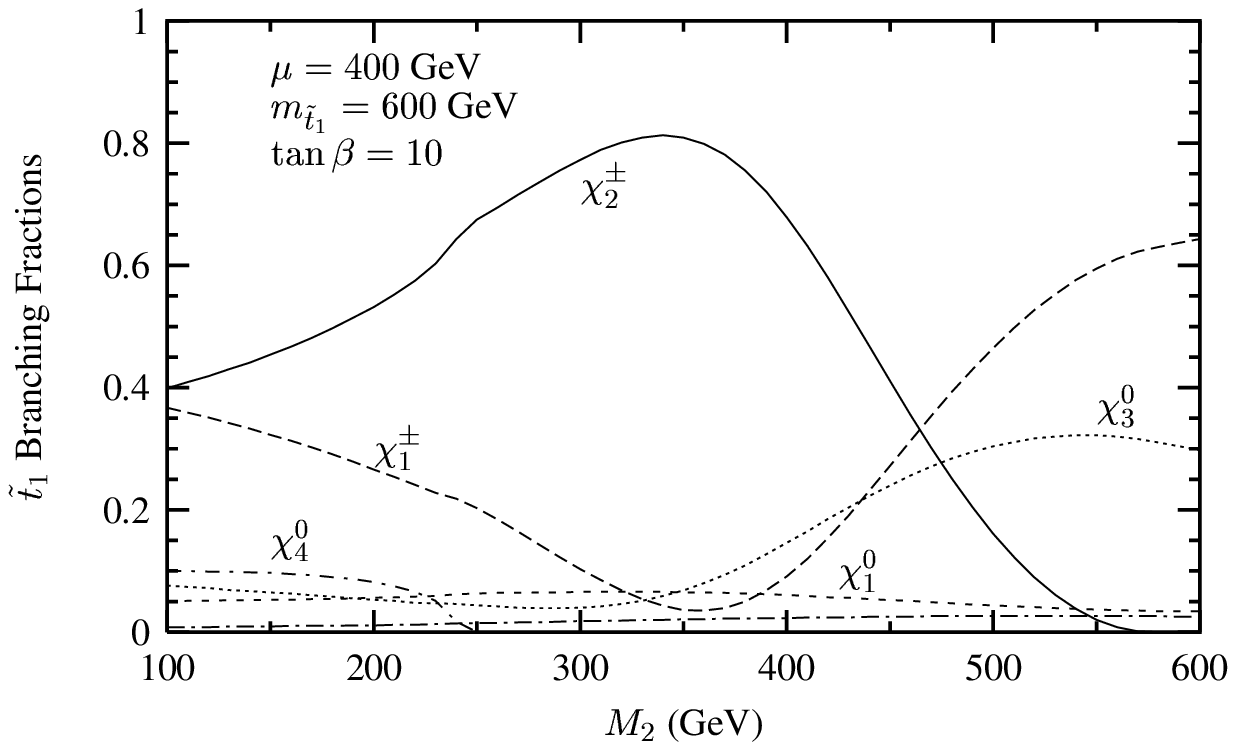,width=23cm}}
\vskip-19cm
\end{center}
\caption{Branching ratios for top squark decaying into charginos and 
neutralinos as a function of $\mu$ with $M_2=150$ GeV (top) and $M_2$ 
with $\mu=400$ GeV (bottom). We have fixed the other parameters to $\tb=10$ 
and  $m_{\tilde{t}_1}=600$ GeV.}
\end{figure}

\subsection*{2.3 Gluino decays}

If gluinos are heavier than squarks, their only relevant decay channel will be 
into quark plus squark final states, $\tilde{g} \to q \tilde{q}$. The partial 
decay width, including the quark mass and the squark mixing angle to take into 
account the possibility of decays into top squarks, using the notation 
introduced previously, is given by [here, $\mu_X=m_X/m_{\tilde{g}}$]:
\beq
\Gamma( \tilde{g} \ra q \tilde{q}_i) &=& \frac{\alpha_s \lambda^{\frac{1}{2}} 
(\mu_{q}^2,\mu_{\tilde{q}_i}^2)}{8} m_{\tilde{g}}
             \bigg[ 1 + \mu_{q}^2 - \mu_{\tilde{q}_i}^2 +
4 a_{i \tilde{g}}^{\tilde{q}} b_{i\tilde{g}}^{\tilde{q}} 
\mu_{q} \mu_{\tilde{g}} \bigg] 
\eeq
In the case, where the gluino is lighter than the squarks, it will mainly decay
into final states involving a quark--antiquark pair and charginos or 
neutralinos. The 
Dalitz density for the decay, taking into account the masses of the final 
fermions, is given by \cite{C2}: 
\beq
\frac{ \dx \Gamma} {\dx x_u \dx x_d} (\tilde{g} \to \chi_j u \bar{d})
&=& \frac{\alpha \alpha_s}{32 \pi} 
m_{\tilde{g}} \,  \sum_{k,l=1}^2 \bigg[ {\rm d}\Gamma^u_{kl} + 
{\rm d}\Gamma^d_{kl} + {\rm d}\Gamma^{ud}_{kl} \bigg]
\eeq
\beq
\dx \Gamma^{\tilde d}_{kl} &=& \frac{1}{(- \mu_u - \mu_{\tilde d_k}+ \hat u)
(- \mu_u - \mu_{\tilde d_l} + \hat u)} \Bigg\{ 
(a^d_{jk} a^d_{jl} + b^d_{jk} b^d_{jl})[ (a^d_{k\tilde{g} } 
a^d_{l\tilde{g}}+ b^d_{k\tilde{g}}b^d_{l\tilde{g}}) ( -\hat u^2 + \hat u  
\non \\ &\times& (1+\mu_\chi+\mu_d+\mu_u) - (\mu_\chi+\mu_d)(1+\mu_u))
-2 \sqrt{\mu_u} (a^d_{k\tilde{g}} b^d_{l\tilde{g}} + a^d_{l\tilde{g}} b^d_{k
\tilde{g}}) (\mu_d+\mu_\chi-\hat u)] 
\non \\
&+& 2(a^d_{jk} b^d_{jl} + a^d_{jl} b^d_{jk}) \sqrt{\mu_\chi \mu_d}
[  (a^d_{k\tilde{g}} a^d_{l\tilde{g}} + b^d_{k\tilde{g}} b^d_{l\tilde{g}})
(\hat{u}-1- \mu_u)-2 \sqrt{\mu_u} (a^d_{k \tilde{g}} b^d_{l\tilde{g}} + 
a^d_{l\tilde{g}} b^d_{k\tilde{g}}) ] \bigg\} \non \\
\dx \Gamma^{\tilde u}_{kl} &=& \frac{1}{(- \mu_d - \mu_{\tilde u_k}+ \hat t)
(- \mu_d - \mu_{\tilde u_l} + \hat t)} \Bigg\{ 
(a^u_{jk} a^u_{jl} + b^u_{jk} b^u_{jl})[ (a^u_{k\tilde{g} } 
a^u_{l\tilde{g}}+ b^u_{k\tilde{g}}b^u_{l\tilde{g}}) ( -\hat t^2 + \hat t  
\non \\ &\times& (1+\mu_\chi+\mu_u+\mu_d) - (\mu_\chi+\mu_u)(1+\mu_d))
-2 \sqrt{\mu_d} (a^u_{k\tilde{g}} b^u_{l\tilde{g}} + a^u_{l\tilde{g}} b^u_{k
\tilde{g}}) (\mu_u+\mu_\chi-\hat t)] 
\non \\
&+& 2(a^u_{jk} b^u_{jl} + a^u_{jl} b^u_{jk}) \sqrt{\mu_\chi \mu_u}
[  (a^u_{k\tilde{g}} a^u_{l\tilde{g}} + b^u_{k\tilde{g}} b^u_{l\tilde{g}})
(\hat{t}-1-\mu_d) -2 \sqrt{\mu_d} (a^u_{k \tilde{g}} b^u_{l\tilde{g}} 
+ a^u_{l\tilde{g}} b^u_{k\tilde{g}}) ] \bigg\} \non \\
\dx \Gamma^{\tilde u\tilde d}_{kl} &=& \sum_{k,l=1}^2 \frac{-2}
{(- \mu_u - \mu_{\tilde d_l} + \hat u)
(- \mu_d - \mu_{\tilde u_k} + \hat t)} \Bigg\{ [a^u_{k \tilde{g}} a^u_{jk} 
b^d_{l\tilde{g}} b^d_{jl} + a^d_{l\tilde{g}} a^d_{jl} b^u_{k\tilde{g}} 
b^u_{jk}] \sqrt{\mu_u\mu_d} \non \\ && (\hat u+\hat t-\mu_u-\mu_d) 
+ [a^u_{k \tilde{g} } a^u_{jk} a^d_{l\tilde{g} } b^d_{jl} + b^u_{k\tilde{g}} 
b^u_{jk} b^d_{l\tilde{g} } a^d_{jl}] \sqrt{\mu_d} (\hat t-\mu_\chi-\mu_u)
\non\\
&& + [a^u_{jk}a^d_{l\tilde{g}} a^d_{jl}b^u_{k\tilde{g}} + b^u_{jk}
b^d_{l\tilde{g}} b^d_{jl}a^u_{k\tilde{g} }]
\sqrt{\mu_\chi\mu_d}(\hat u-\mu_u-1) -2[a^u_{jk} a^d_{jl} b^u_{k\tilde{g}} 
b^d_{l\tilde{g}} + a^u_{k\tilde{g}} a^d_{l\tilde{g}} b^u_{jk} b^d_{jl}] 
\non\\
&& \sqrt{\mu_\chi} \sqrt{\mu_u\mu_d} +
[a^u_{jk} b^u_{k\tilde{g}} b^d_{l\tilde{g}} b^d_{jl} 
+ a^u_{k\tilde{g}} a^d_{l\tilde{g}} a^d_{jl} b^u_{jk}] \sqrt{\mu_u} 
(\hat u-\mu_\chi-\mu_d) \non\\
&& + [a^u_{jk} a^d_{l\tilde{g}} b^u_{k\tilde{g}} b^d_{jl} + a^u_{k\tilde{g}} 
a^d_{jl}  b^u_{jk} b^d_{l\tilde{g}}]
(\hat u \hat t -\mu_\chi -\mu_u \mu_d) +
[a^u_{k\tilde{g}} a^u_{jk} a^d_{l\tilde{g}} a^d_{jl} + b^u_{k\tilde{g}}
b^u_{jk} b^d_{l\tilde{g}} b^d_{jl}] 
\sqrt{\mu_\chi} 
\non\\
&& (\hat u+\hat t-\mu_\chi-1) + [a^u_{k\tilde{g}} a^u_{jk} a^d_{jl} 
b^d_{l\tilde{g}} 
+ b^u_{k\tilde{g}}b^u_{jk} b^d_{jl} a^d_{l\tilde{g}}] \sqrt{\mu_\chi\mu_u} 
(\hat t-\mu_d-1) \Bigg\}
\eeq
where $x_u=2E_u/m_{\tilde g }, x_d= 2E_{d}/m_{\tilde g}$ are the reduced 
energies of the final quarks, $\mu_X= m_X^2/m^2_{\tilde g}$ the reduced 
masses and $\hat{t}=(p _{\tilde{g}} - p_u)^2/m^2_{\tilde{g} }= 1- x_u+\mu_u$,
$\hat{u}=(p _{\tilde{g}} - p_d)^2/m^2_{\tilde{g} }= 1- x_d+\mu_d$. 
The squark-quark couplings to charginos and neutralinos,  $a^q_{jl}$ and 
$b^q_{jl}$ and the couplings to gluinos  $a^q_{l \tilde{g}}$ and 
$b^q_{l\tilde{g}}$ have been given previously.
The fully integrated partial decay width, in the case of massless final state
quarks, can be found in Ref.~\cite{C3}. \smallskip

\begin{figure}[htbp]
\begin{center}
\vskip-5cm
\hskip-1cm\centerline{\epsfig{file=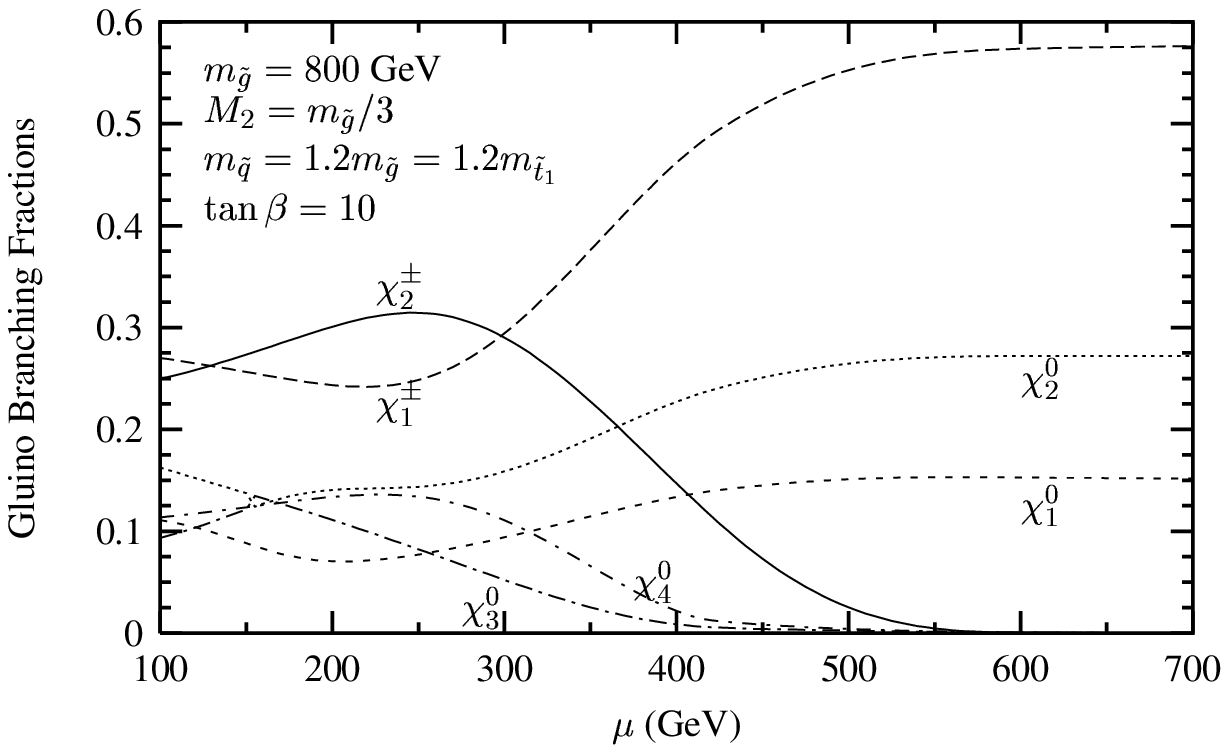,width=23cm}}
\vskip-24cm
\hskip-1cm\centerline{\epsfig{file=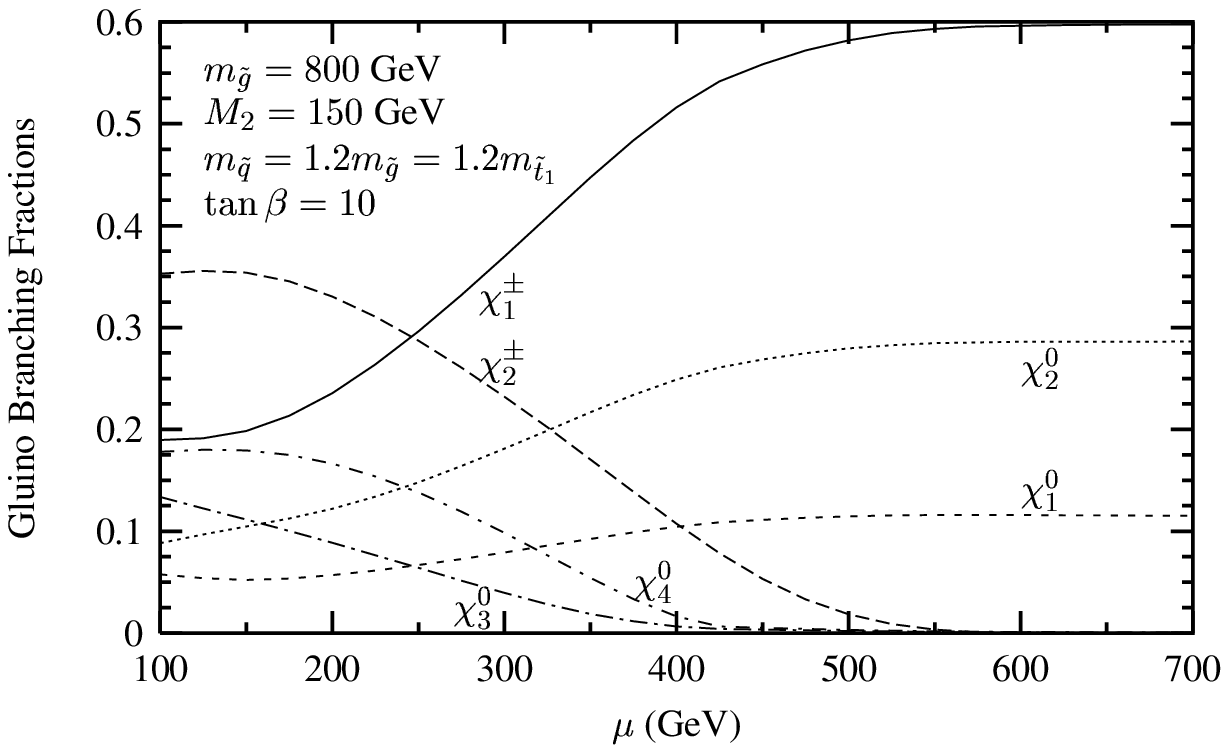,width=23cm}}
\vskip-19cm
\end{center}
\caption{Branching ratios of gluinos into electroweak gauginos as functions of 
$\mu$ for $\tb=10$. The gluino mass is fixed to $m_{\tilde{g}}=800$ GeV while 
the squark masses are  $m_{\tilde{q}} = 1.2 m_{\tilde{t}_1} =1.2 
m_{\tilde{g}}$. The wino mass parameter is fixed to $M_2=2 M_1= 
m_{\tilde{g}}/3$ (upper curve) and $M_2=2M_1=150$ GeV (lower curve).}
\end{figure}

The various gluino branching fractions are shown in Fig.~4 as a function of 
$\mu$ for a gluino mass $m_{\tilde{g}} =800$ GeV and for squark masses fixed to
$m_{\tilde{q}}=1.2 m_{\tilde{t}_1} = 1.2 m_{\tilde{g}}$. Two different
scenarii are exemplified: in the upper panel of Fig.~4, the approximate GUT
relation $M_2 = m_{\tilde{g}}/3$ is adopted while in the lower part of Fig.~4,
this GUT relation is relaxed and we set $M_2=150$ GeV, i.e. with gaugino--like
$\chi_1^\pm, \chi_{1,2}^0$ for large $\mu$ values. [We however, keep the 
equality of the wino and bino mass parameters at the GUT scale, leading
to $M_2\simeq 2M_1$ at low energies].  In both scenarii, the top
squark is lighter compared to other squarks [which is generally the case, as
discussed previously], $m_{\tilde{t}_1} = m_{\tilde{g}}$, and its virtuality in
the decay of the gluinos is smaller. Since, in addition, the top squark has
large couplings to the higgsino states, the decay modes $\tilde{g} \to t \bar{b}
\chi_{i}^-$ and $\tilde{g} \to t \bar{t} \chi_i^0$ [and the charge conjugate
states] are more important, the former being in general dominant because of the
larger phase--space and the stronger charged current couplings. \s 

In the universal scenario with $M_2 \sim m_{\tilde{g}}/3 \sim 270$ GeV, the
branching ratio for the decay $\tilde{g} \to t\bar{b} \chi_2^-$ is at the level
of $\sim 30\%$ for small values of $\mu \lsim 300$ GeV, where $\chi_{1,2}^\pm$
are mixtures of gauginos and higgsinos, while the branching ratios of the
decays into top quarks and $\chi_{3,4}^0$ final states are smaller by a factor
of 3.  The branching ratios decrease with increasing $\mu$ since $\chi_{3,4}^0$
and $\chi_2^\pm$ become heavier and the phase space for the decay is reduced. 
However, even for values of $\mu$ around 400 GeV, BR($\tilde{g} \to t
\bar{b}\chi_2^-$) stays above the level of 10\%.  In the non--universal
scenario with $M_2=150$ GeV, $\chi_2^\pm$ are almost higgsino--like for
relatively large $\mu$ values, $\mu \gsim 300$ GeV.  BR($\tilde{g} \to b t
\chi_2^-$) is larger than what was previously for small $\mu$ values while for
large $\mu$ values, it is similar to the previous case.  

\section*{3. Decays into Charged Higgs Bosons}

\subsection*{3.1 Two--body decays of heavy charginos and neutralinos}

The heavier chargino and neutralinos will mainly decay into the lighter 
chargino and neutralino states and, if enough phase space is available, into
sfermion--fermion pairs. The partial decay widths of the two--body decays,
[including the possibility of massive fermions in the last case], are given
by:
\beq
\Gamma(\chi_i \ra f  \tilde{f}_j) = \frac{\alpha N_c}{8}\, m_{\chi_i} \, \bigg[
 \left( (a_{ij}^f)^2 + (b_{ij}^f)^2 \right) (1- \mu_{\tilde f_j} + \mu_f)
+ 4 \sqrt{\mu_f} a_{ij}^f b_{ij}^f \bigg] \, \lambda^{\frac{1}{2}} (\mu_f, 
\mu_{\tilde f_j})
\eeq
\beq
\Gamma(\chi_i \ra \chi_j V) = \frac{\alpha}{8} \, m_{\chi_i} \, 
\lambda^{\frac{1}{2}}(\mu_{\chi_j},\mu_V) \left\{ -12 \sqrt{\mu_{\chi_j}}
G_{jiV}^L G_{jiV}^R \right. \hspace*{3.8cm} \non \\
\hspace*{.9cm} + \left. \left[ (G_{jiV}^L)^2 + (G_{jiV}^R)^2 \right] 
(1+ \mu_{\chi_j}-\mu_V) +(1- \mu_{\chi_j} +\mu_V)(1- \mu_{\chi_j}-\mu_V) 
\mu_V^{-1} \right\}
\eeq
\beq
\Gamma(\chi_i \ra \chi_j H_k) = \frac{\alpha}{8} \, m_{\chi_i} \, 
\lambda^{\frac{1}{2}}(\mu_{\chi_j},\mu_{H_k}) \left\{ \left[ (G^L_{ijk})^2
+(G^R_{ijk})^2 \right] ( 1+ \mu_{\chi_j} -\mu_{H_k}) \right. \non \\
+\left. 4 \sqrt{\mu_{\chi_j}} \, G^L_{ijk} G^R_{ijk}  \right\} \hspace*{7cm}
\eeq
The couplings among charginos, neutralinos and fermions/sfermions have been
given previously, while the chargino and neutralino couplings to the Higgs and
the gauge bosons are given by [$H_k=h,H,A$ and $H^\pm$ for $k=1,2,3,4$]: 
\beq
G^{L,R}_{\chi^0_i \chi^+_j W^+} = G^{L,R}_{ijW} & {\rm with} & 
\begin{array}{l} 
G^L_{ijW} =  \frac{1}{\sqrt{2}s_W} [-Z_{i4} V_{j2}+\sqrt{2}Z_{i2} V_{j1}] \\
G^R_{ijW} =  \frac{1}{\sqrt{2}s_W} [Z_{i3} U_{j2}+ \sqrt{2} Z_{i2} U_{j1}] 
\end{array}  
\eeq
\beq
G^{L,R}_{\chi^-_i \chi^+_j Z} = G^{L,R}_{ijZ} & {\rm with} & 
\begin{array}{l} 
G^L_{ijZ} = \frac{1}{c_W s_W} \left[- \frac{1}{2} V_{i2} V_{j2} - 
V_{i1} V_{j1} +\delta_{ij}s_W^2 \right]  \\
G^R_{ijZ} = \frac{1}{c_W s_W} \left[- \frac{1}{2} U_{i2} U_{j2} - 
U_{i1} U_{j1} +\delta_{ij}s_W^2 \right] 
\end{array} 
\eeq
\beq
G^{L,R}_{\chi^0_i \chi^0_j Z} = G^{L,R}_{ijZ} & {\rm with} &
\begin{array}{l} 
G^L_{ijZ} = - \frac{1}{2s_Wc_W} [Z_{i3} Z_{j3} - Z_{i4} Z_{j4}]  \\
G^R_{ijZ} = + \frac{1}{2s_Wc_W} [Z_{i3} Z_{j3} - Z_{i4} Z_{j4} ] 
\end{array} 
\eeq
\beq
G^{L,R}_{\chi^0_i \chi^+_j H^+} = G^{L,R}_{ij4} & {\rm with} &  
\begin{array}{l} 
G^L_{ij4} = \frac{c_\beta}{s_W} \left[ Z_{j4} V_{i1} + \frac{1}{\sqrt{2}} 
\left( Z_{j2} + \tan \theta_W Z_{j1} \right) V_{i2} \right] \non \\
G^R_{ij4} = \frac{s_\beta}{s_W} \left[ Z_{j3} U_{i1} - \frac{1}{\sqrt{2}}
\left( Z_{j2} + \tan \theta_W Z_{j1} \right) U_{i2} \right]
\end{array} 
\eeq
\beq
G^{L,R}_{\chi^-_i \chi^+_j H^0_k} = G^{L,R}_{ijk} & {\rm with} & 
\begin{array}{l} 
G^L_{ijk}= \frac{1}{\sqrt{2}s_W} \left[ e_k V_{j1}U_{i2}-d_k V_{j2}U_{i1}
\right] \\
G^R_{ijk}= \frac{1}{\sqrt{2}s_W} \left[ e_k V_{i1}U_{j2}-d_k V_{i2}U_{j1}
\right] \epsilon_k 
\end{array} 
\eeq
\beq
G^{L,R}_{\chi^0_i \chi^0_j H_k} = G^{L,R}_{ijk} & {\rm with} & 
\begin{array}{l} 
G^L_{ijk} = \frac{1}{2 s_W} \left( Z_{j2}- \tan\theta_W Z_{j1} \right) 
\left(e_k Z_{i3} + d_kZ_{i4} \right) \ + \ i \leftrightarrow j
 \\
G^R_{ijk} = \frac{1}{2 s_W}  \left( Z_{j2}- \tan\theta_W Z_{j1} 
\right) \left(e_k Z_{i3} + d_kZ_{i4} \right) \epsilon_k \ + \ i 
\leftrightarrow j
\end{array} 
\eeq
where $\epsilon_{1,2}=- \epsilon_3 =1$ and the coefficients $e_k$ and $d_k$ 
read
\begin{eqnarray}
e_1/d_1=c_\alpha/  -s_\alpha \ , \
e_2/d_2=-s_\alpha / -c_\alpha \ ,  \
e_3/d_3=-s_\beta / c_\beta
\end{eqnarray} 

The branching ratios of the heavier chargino $\chi_2^\pm$ and $\chi_{3,4}^0$
into the lighter ones $\chi_1^\pm$ and $\chi_{1,2}^0$ and gauge and Higgs
bosons are shown in Figs.~5 and 6 for, respectively, the two scenarii (i)
gaugino--limit and (ii) higgsino--limit discussed previously. Squarks and
sleptons are assumed to be too heavy to play a role here\footnote{We assume in
our analysis that the heavier charginos and neutralinos are coming from the
decays of heavier squarks, including top squarks which is dealt as a special
case. Decays of charginos  and neutralinos into sleptons, which can be lighter
than squarks, are relevant only if the former particles are gaugino--like since
the higgsino--slepton--lepton couplings are rather tiny, unless $\tb$ is very
large in which case the decays into $\tilde{\tau}$'s could play a role.
However, in the large $\tb$  case, $H^\pm$ particles can be produced directly
with large cross sections.}. To analyze them, it is useful to discuss first the
couplings of charginos and neutralinos to the Higgs and gauge bosons.\s

The Higgs bosons couples preferentially to mixtures of gauginos and higgsinos. 
This means that in the gaugino--like or higgsino like regions, the couplings of
the Higgs bosons which involve heavy and light chargino/neutralino states are
maximal, while the couplings involving only heavy or light ino states are
suppressed by powers of $M_2/\mu$ for $|\mu| \gg M_2$ or powers of $\mu/M_2$
for $|\mu| \ll M_2$.  To the contrary, the gauge boson couplings to charginos
and neutralinos are important only for higgsino--like states. Thus, in
principle, the (higgsino or gaugino--like) heavier chargino and neutralinos
$\chi_2^\pm$ and $\chi_{3,4}^0$, will dominantly decay, if phase space allowed,
into Higgs bosons and the lighter $\chi$ states. \s
 
However, in the asymptotic limit where the heavier chargino and neutralino 
masses are very large, $m_{\chi_i} \gg m_{\chi_j}, M_{H_k}, M_V$, the decay 
widths into Higgs bosons grow as $m_{\chi_i}$, 
\beq
\Gamma(\chi_i \ra \chi_j H_k) \sim \frac{1}{8} \alpha m_{\chi_i} [ 
(G^L_{ijk})^2 +(G^R_{ijk})^2 ]
\eeq
while the decay widths into gauge bosons grow as $m_{\chi_i}^3$
\beq
\Gamma(\chi_i \ra \chi_j V) \sim \frac{\alpha m^3_{\chi_i}} {8 M_V^2} [ 
(G^L_{ijV})^2 +(G^R_{ijV})^2 ]
\eeq
This is due to the longitudinal component of the gauge boson propagators which
introduce extra powers of the $\chi_i$ four--momentum in the decay amplitudes. 
The suppression of the $(G^{L,R}_{ijV})^2$ squared coupling by powers of
$(\mu/M_2) ^2$ or $(M_2/\mu)^2$ depending on whether we are in the gaugino or
higgsino region will be compensated by the power $m_{\chi}^2/M_Z^2$ from the
matrix element squared. Therefore, the branching ratios for the decays of heavy
$\chi$ particles into lighter ones and Higgs or gauge bosons will have the same
order of magnitude. Of course, as usual, the charged current decay modes will
be more important than the neutral current decay modes. \s

This is exemplified in the figures. In both higgsino and gaugino regions,
Fig.~5 and Fig.~6 respectively, the decays of charginos $\chi_2^\pm$ and 
neutralinos $\chi_{3,4}^0$ into lighter charginos and neutralinos and 
Higgs bosons are not the dominant ones. Still, decays into $H^\pm$ bosons 
will have substantial branching fractions of the order of 20 to 30\%. 

\begin{figure}[htbp]
\begin{center}
\vskip-4.cm
\hskip-1cm\centerline{\epsfig{file=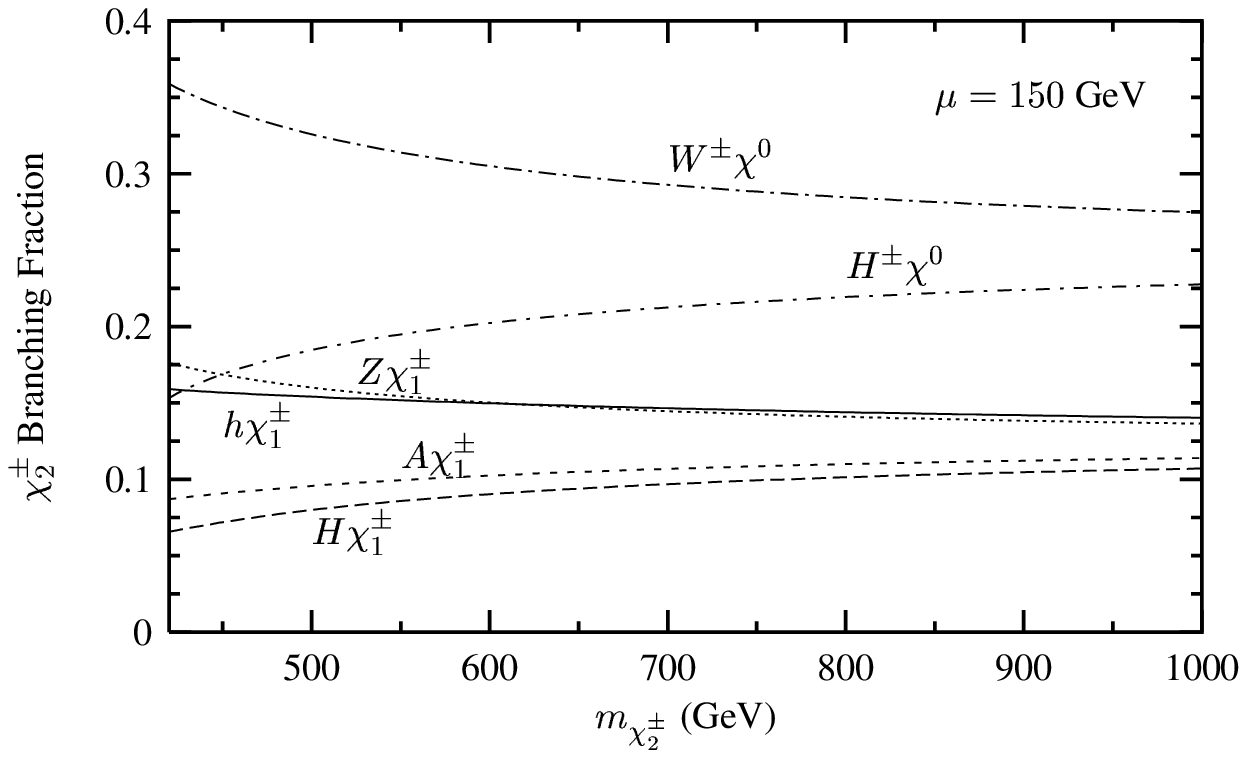,width=17cm}}
\vskip-17.5cm
\hskip-1cm\centerline{\epsfig{file=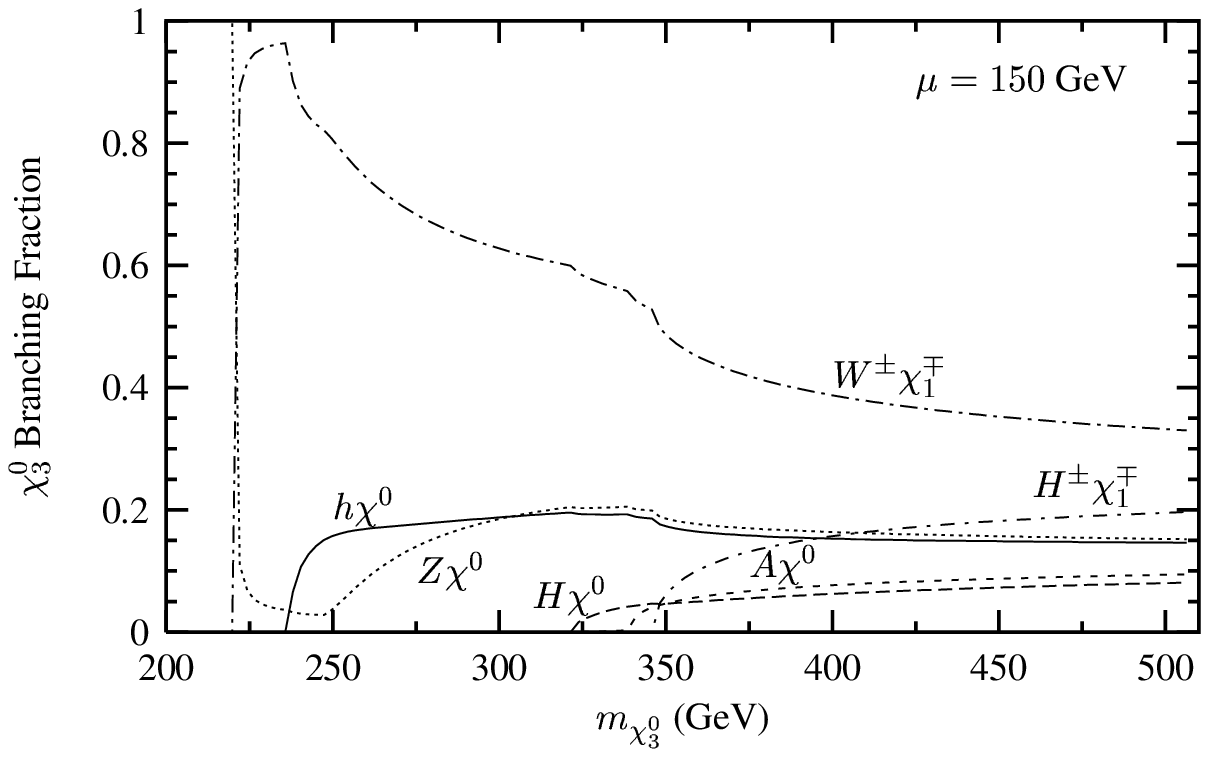,width=17cm}}
\vskip-17.5cm
\hskip-1cm\centerline{\epsfig{file=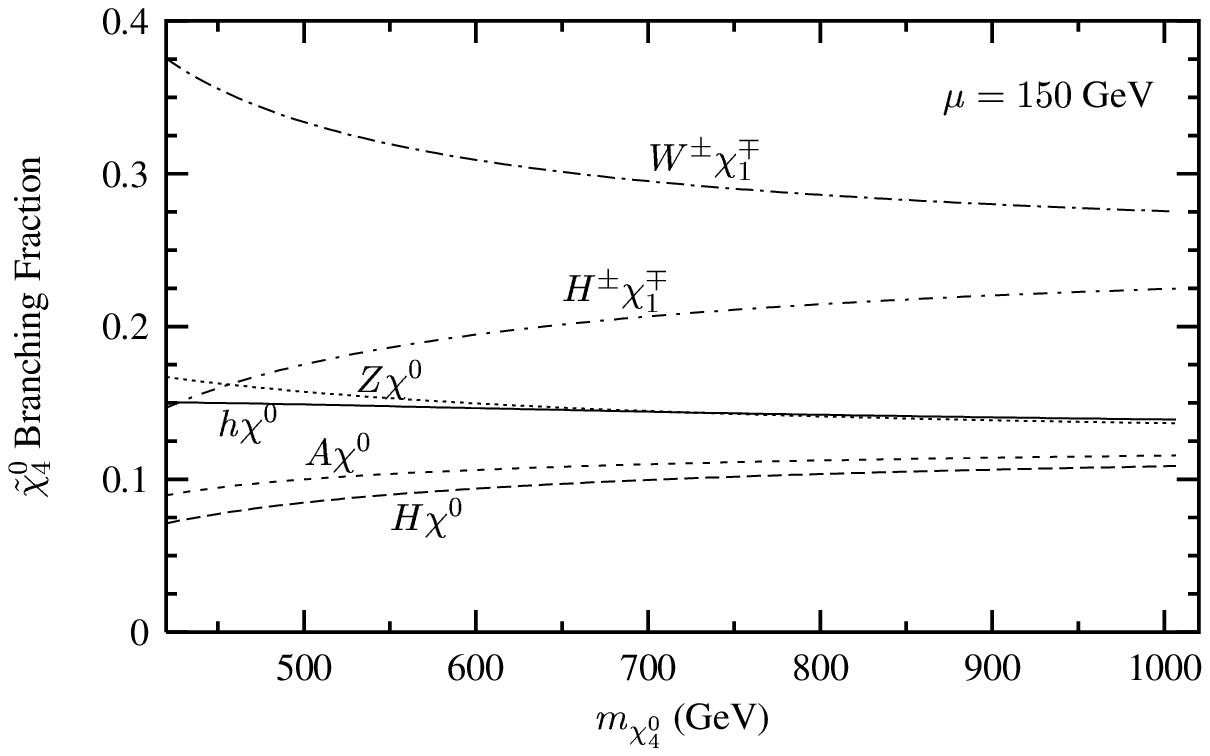,width=17cm}}
\vskip-14cm
\caption{Branching ratios of heavier chargino and neutralinos into the 
lighter ones and gauge/Higgs bosons as functions of their masses for $\tb=10$. 
The charged Higgs boson mass is $M_{H^\pm}=200$ GeV, $\mu$ is fixed to 150 GeV 
while $M_2$ varies with the heavy ino mass; $\chi^0$ represents the lighter 
$\chi_1^0$ and $\chi_2^0$ neutralinos for which the branching ratios are added.}
\end{center}
\end{figure}

\begin{figure}[htbp]
\begin{center}
\vskip-4.cm
\hskip-1cm\centerline{\epsfig{file=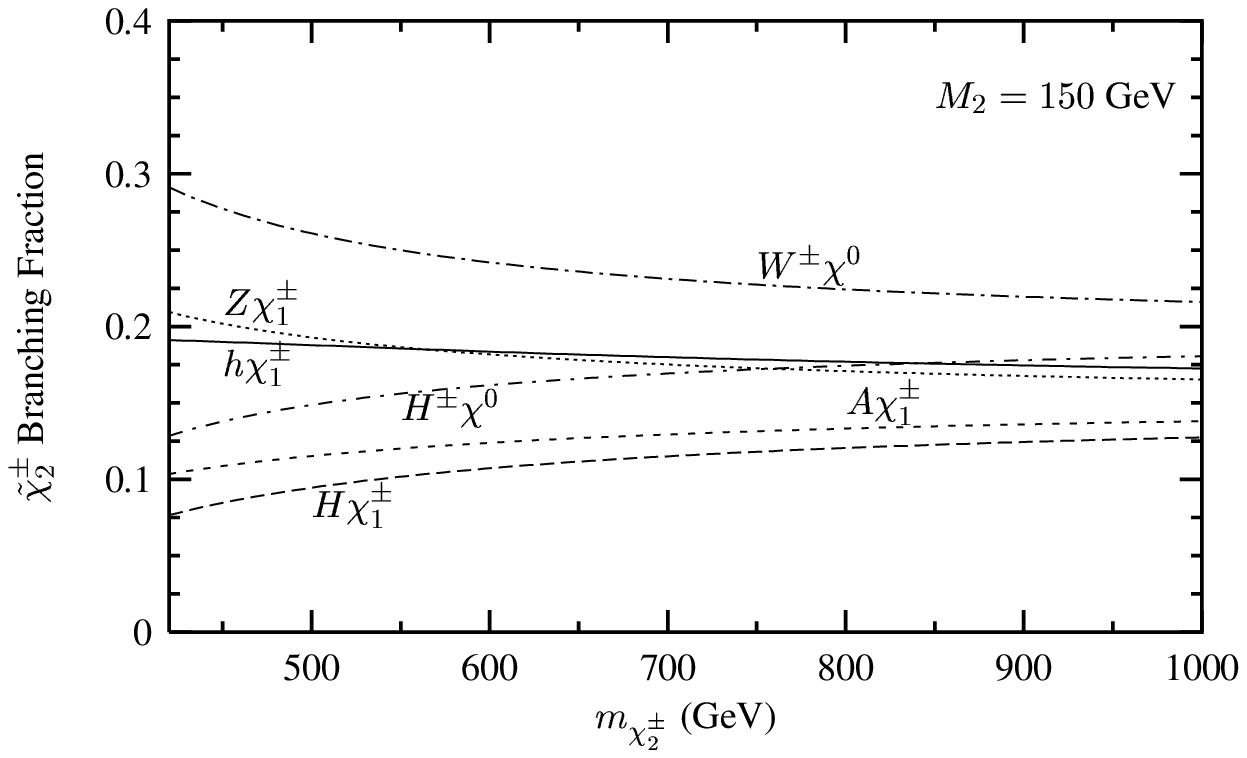,width=17cm}}
\vskip-17.5cm
\hskip-1cm\centerline{\epsfig{file=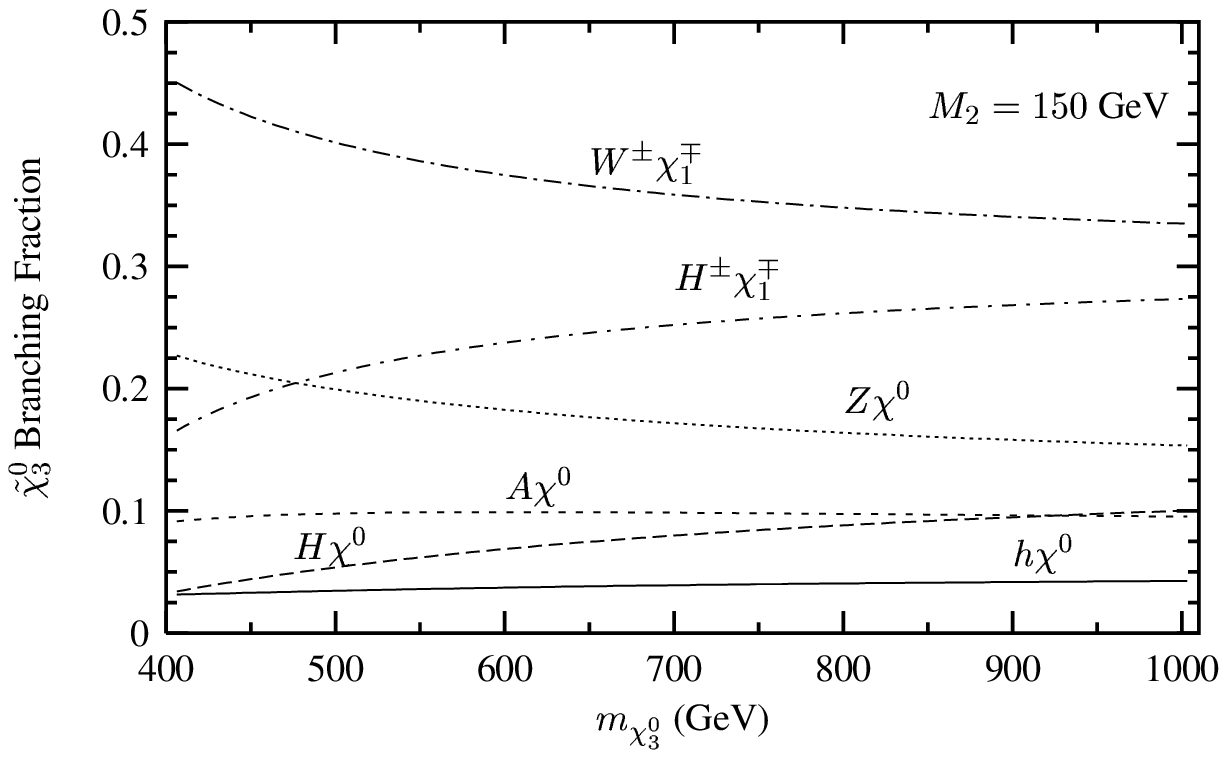,width=17cm}}
\vskip-17.5cm
\hskip-1cm\centerline{\epsfig{file=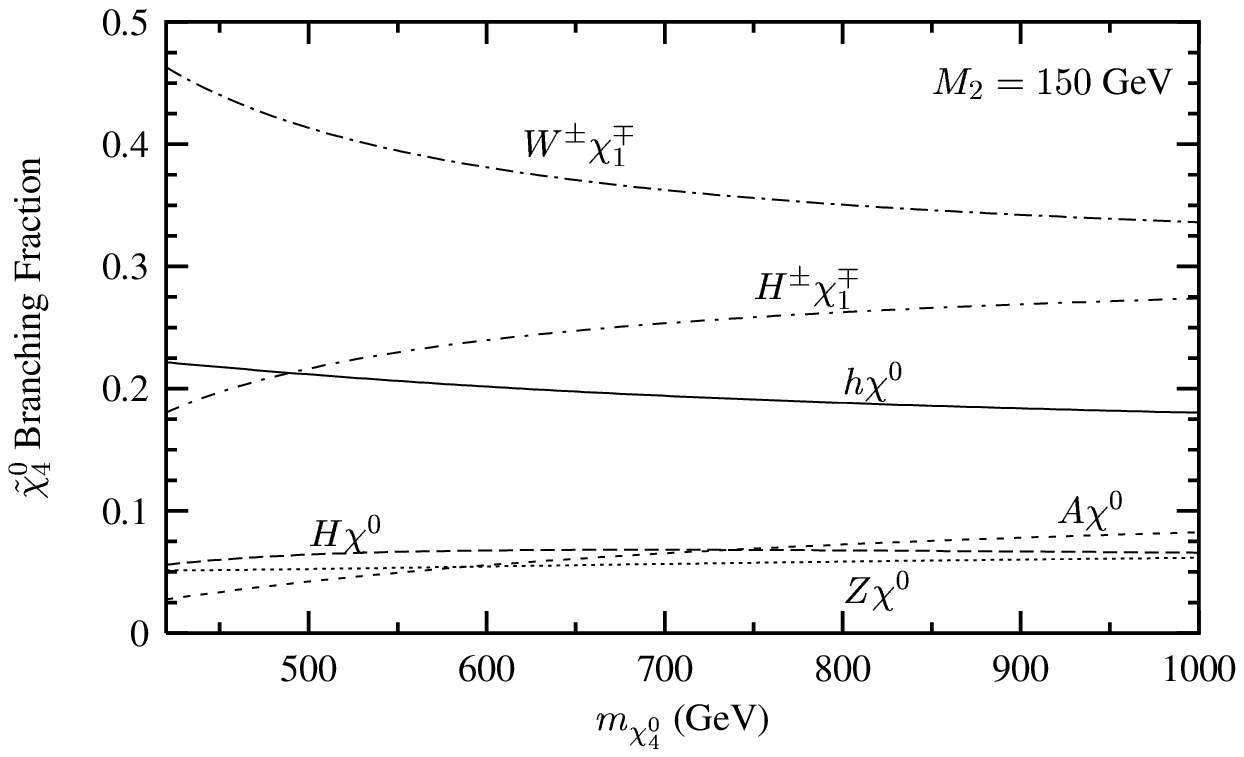,width=17cm}}
\vskip-14cm
\caption{Branching ratios of heavier chargino and neutralinos into the 
lighter ones and gauge/Higgs bosons as functions of their masses for $\tb=10$. 
The charged Higgs boson mass is $M_{H^\pm}=200$ GeV, $M_2$ is fixed to 150 GeV 
while $\mu$ varies with the heavy ino mass; $\chi^0$ represents the lighter 
$\chi_1^0$ and $\chi_2^0$ neutralinos for which the branching ratios are added.}
\end{center}
\end{figure}

\subsection*{3.2 Cascade decays of squarks and gluinos into $H^\pm$ bosons}

\subsubsection*{3.2.1 The case of squarks}

We first focus on the scenario where gluinos are heavier than the scalar
partners of the light quarks and in this case, all gluinos produced in the
processes $pp \to \tilde{g} \tilde{g}$ and $pp \to \tilde{q} \tilde{g}$ will
decay into squarks and light quarks. The squarks will subsequently decay into
their partner quarks and all types of charginos or neutralinos. The heavier
charginos and neutralinos will then decay into gauge bosons and Higgs bosons
and the lighter chargino and neutralino states. Fig.~7 shows the final cross
sections times branching ratio to obtain {\it one} charged Higgs boson in the
final state, $\sigma \times {\rm BR}(\to H^\pm)$. For the total production 
cross sections, we use the same set of parton densities and the same scale as 
in section 2.1. \s

The particularly favorable case that we will discuss here is as follows. We
choose a common squark mass of $m_{\tilde{q}} =800$ GeV and fix $\tb=10$. We
also fix $\mu=150$ GeV and vary the wino mass parameter $M_2$ in such a way
that for $M_2 \gsim 300$ GeV we have higgsino--like lighter chargino and
neutralinos.  For the gluino mass, we will assume in the first step that the
gaugino masses are unified at the GUT scale so that $m_{\tilde{g}}= 3M_2$, and
in the second step, we relax this universality assumption and fix the gluino
mass to a constant value of $m_{\tilde{g}}=900$ GeV, while still varying $M_2$.
In the latter scenario, the gluino is not allowed kinematically to decay into
top squarks if $m_{\tilde{t}_1} \simeq m_{\tilde{g}}$, but in the universal
case since $m_{\tilde{g}} \sim 3 M_2 \gsim m_{\tilde{t}_1}+m_t$, gluino decays
into stops have to be taken into account.  In the analysis, we will treat all
squarks on equal footing and assume that the branching fraction for top squarks
decaying into the heavier charginos and neutralinos are the same as for the
scalar partners of light squarks. [This is a rather conservative approach since
we have seen in the previous section that the branching ratios for stop decays
into bottom quarks and the heavier chargino states can be larger.] The charged
Higgs bosons masses are chosen to be $M_{H^\pm} =180$, 200 and 300 GeV. \s 

In the universal scenario of Fig.~7 [upper panel], we see that the cross
sections times branching ratios  for $H^\pm$ final states exceed the level of
0.1 pb in most of the displayed $M_2$ range. This means that with a luminosity
of $\int {\cal L}=300$ fb$^{-1}$ which is expected to be collected at the LHC,
around $\sim 30.000$ charged Higgs bosons can be produced through the cascade
decays in this scenario. For small $M_2$ values, the states $\chi_{3,4}^0$ and
$\chi_{2}^+$ are not heavy enough for the decays into $H^\pm$ bosons to occur,
in particular for large $M_{H^\pm}$. When these decays are allowed, $\sigma
\times {\rm BR}(\to H^\pm)$ values of the order of 1 pb for $M_{H^\pm} \sim
180$ GeV and 0.3 pb for $M_{H^\pm} \sim 300$ GeV can be reached.  For
increasing values of $M_2$, the gluino mass increases and the cross section for
associated squark and gluino production, which is the largest in this case, as
well as for gluino pair production drop and $\sigma \times {\rm BR}(\to H^\pm)$
decreases accordingly; at some stage, only the cross section for squark
production survives [since $m_{\tilde{q}}$ is fixed]. The decrease of $\sigma
\times {\rm BR}(\to H^\pm)$ with increasing $M_2$ is also due to the more
suppressed phase space for $\tilde{q} \to q' \chi_2^\pm, q\chi_4^0$ since for
large $M_2$, $m_{\chi_4^0} , m_{\chi_2^\pm} \sim M_2$. For even larger $M_2$
values, $M_2 \gsim 650$ GeV, the channel $\chi_3^0 \to H^\pm \chi_1^\mp$ opens
up, and since the phase space is more favorable, because $m_{\chi_3^0} 
\sim M_2/2$, $\sigma \times {\rm BR}(\to H^\pm)$ increases again. \s

In the non--universal case [lower panel of Fig.~7], with a gluino mass fixed to
$m_{\tilde{g}}=900$ GeV, the cross sections times branching ratios for $H^\pm$
final states can be much larger. In the low $M_2$ region, the situation is
similar to the previous case in which the gluino was rather light and
contributed substantially to the production cross sections.  However, for
increasing $M_2$ values, $\sigma_{\tilde{g} \tilde{g}}$ and $\sigma_{\tilde{g}
\tilde{q}}$ stay constant contrary to the universal case, while the phase space
for the decays of the heavier charginos and neutralinos into charged Higgs
bosons increases. $\sigma \times {\rm BR}(\to H^\pm)$ can then reach the
picobarn level even for relatively large charged Higgs boson masses,
$M_{H^\pm}\sim 300$ GeV, leading to samples containing several hundred thousand
charged Higgs bosons events with the high luminosity $\int {\cal L}=300$
fb$^{-1}$.  

\begin{figure}[htbp]
\begin{center}
\vskip-5cm
\hskip-1cm\centerline{\epsfig{file=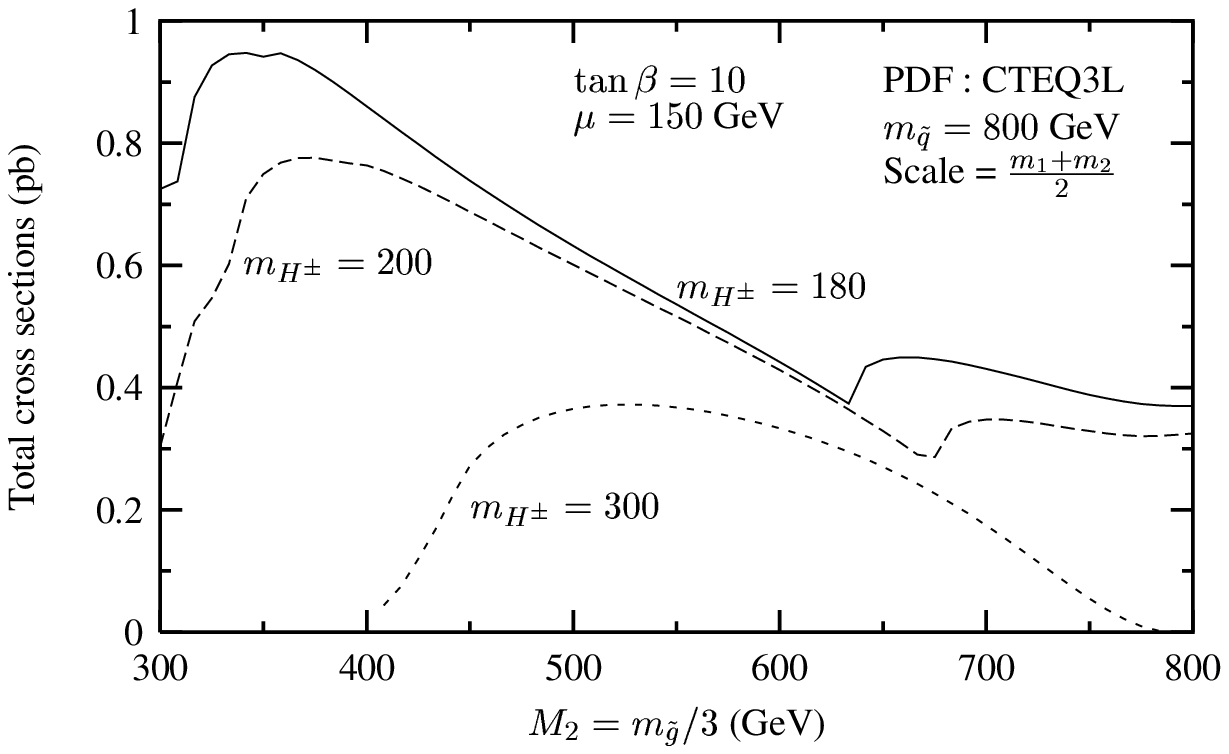,width=23cm}}
\vskip-24cm
\hskip-1cm\centerline{\epsfig{file=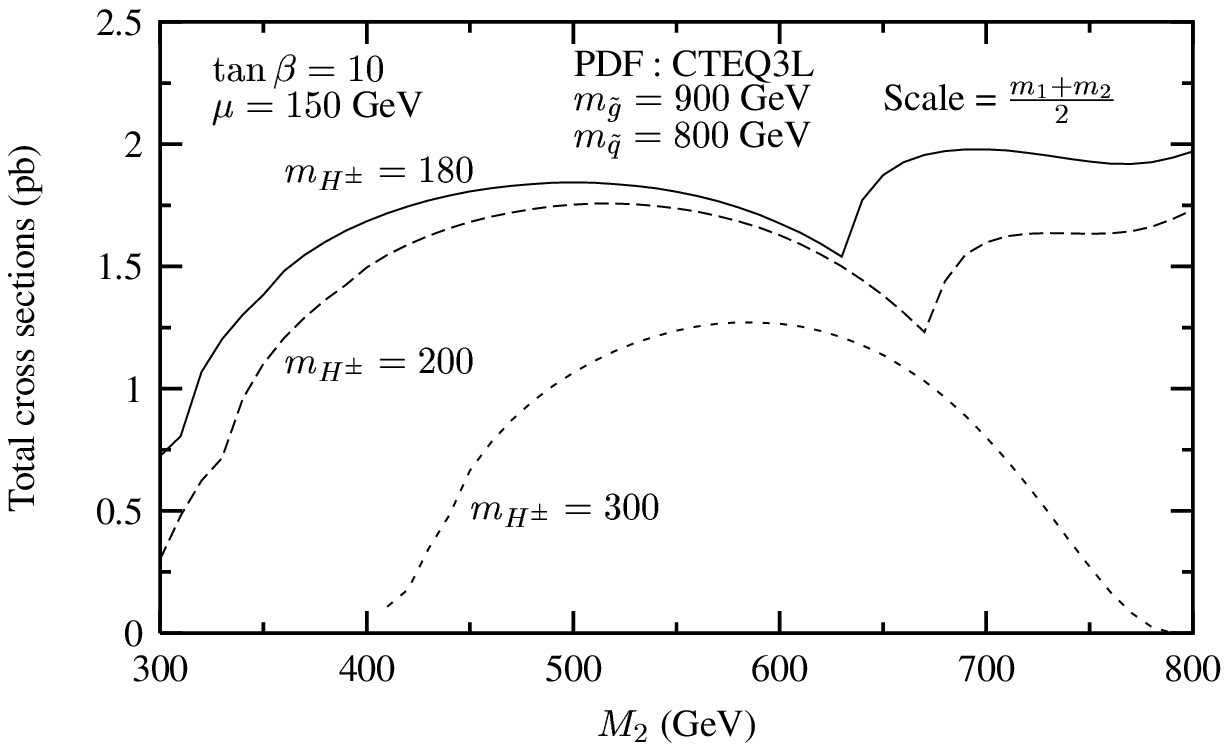,width=23cm}}
\vskip-19cm
\end{center}
\caption{Cross sections times branching ratios for gluinos decaying into 
squarks and squarks decaying through cascades into charged Higgs bosons
with masses $M_{H^\pm}=180,200$ and 300 GeV. They are shown as functions of 
$M_2$ with $\mu=150$ GeV and $\tb=10$. The squark mass is fixed to 
$m_{\tilde{q}}=800$ GeV, while the gluino mass is $m_{\tilde{g}} =3M_2$
(upper curve) and $m_{\tilde{g}}=900$ GeV (lower curve).}
\end{figure}

\subsubsection*{3.2.2 The case of gluinos}

The production cross sections times branching ratios for squarks decaying into
gluinos, with the subsequent three--body decays of the gluinos into the heavier
charginos and neutralinos which then decay into the lighter $\chi$ states and
charged Higgs bosons are shown in Fig.~8. Again we set $\tb=10$ and fix the
gluino and squark masses to $m_{\tilde{q}} =1.2 m_{\tilde{g}}, m_{\tilde{g}}
=800$ GeV; the $H^\pm$ boson masses are taken to be $M_{H^\pm}=180,200$ and 300
GeV.  Then we vary the higgsino parameter $\mu$ and take the universal scenario
where $M_2$ and $M_3$ are related, as well as the non--universal scenario where
$M_2=150$ GeV so that the heavier $\chi$ states are higgsino--like for large
enough values of $|\mu|$.  In both cases the cross sections for squark and
gluino production are constant and the variation of $\sigma \times {\rm
BR}(H^\pm)$ is only due to the variation of the branching ratios BR$(\tilde{g}
\to \chi_{3,4}^0 qq, \chi_2^\pm qq')$ and   BR$(\chi_{3,4}^0, \chi_2^\pm \to
\chi_{1}^\pm H^\mp, \chi_{1,2}^0 H^\pm)$. \s

\begin{figure}[htbp]
\begin{center}
\vskip-5cm
\hskip-1cm\centerline{\epsfig{file=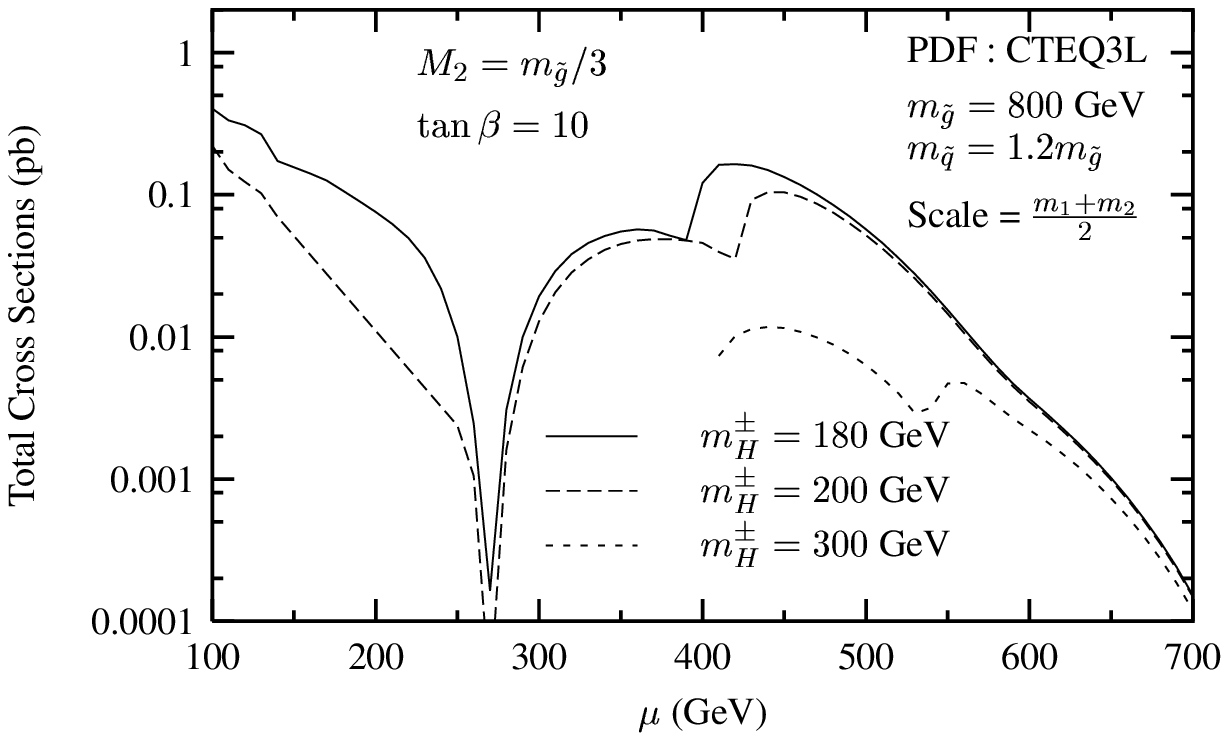,width=23cm}}
\vskip-24cm
\hskip-1cm\centerline{\epsfig{file=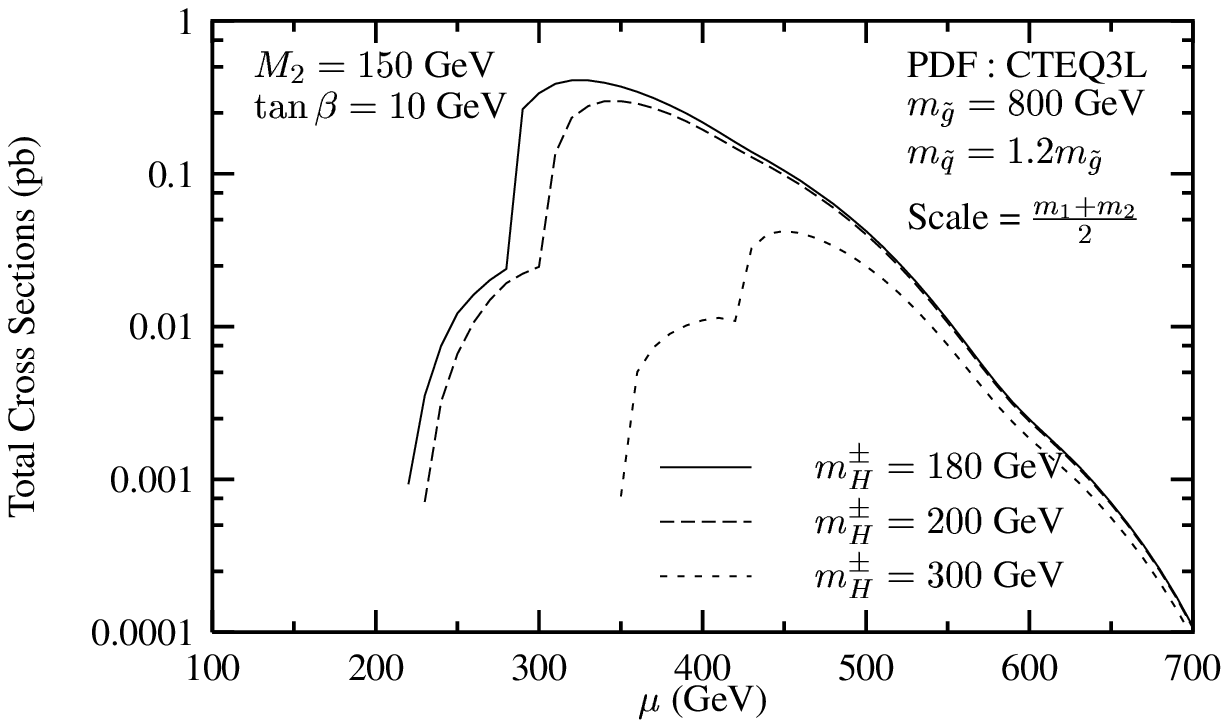,width=23cm}}
\vskip-19cm
\end{center}
\caption{Cross sections times branching ratios for squarks decaying into 
gluinos with the gluinos decaying through cascades into charged Higgs bosons
with masses $M_{H^\pm}=180,200$ and 300 GeV. They are shown as functions of 
$\mu$ for $\tb=10$. The gluino  mass is fixed to $m_{\tilde{g}}=800$ GeV and
the squark mass to $m_{\tilde{q}}=1.2m_{\tilde{g}}$. In the upper figure, 
$M_2=m_{\tilde{g}}/3$ while $M_2=150$ GeV in the lower figure.}
\end{figure}

In the universal scenario, with  $m_{\tilde{g}} \sim M_3 \sim 3M_2$, $\sigma
\times {\rm BR}(H^\pm)$ is relatively large for small values of $\mu$ and
$M_{H^\pm}$, when the gaugino--like heavy $\chi$ states are light enough for
the decays $\chi_{4}^0 \to \chi_1^\pm H^\mp$ and $\chi_{2}^\pm \to \chi_{1,2}^0
H^\pm$ to occur.  In the mixed region, $\mu \sim M_2$, the mass difference
between the heavy and light $\chi$ states are too small to allow for decays in
$H^\pm$ bosons. For large values of $\mu$, $\sigma \times {\rm BR}(H^\pm)$
increases to reach values of the order of $\sim 0.1$ pb for $M_{H^\pm} \sim
200$ GeV [in particular when the additional channels $\chi_3^0 \to \chi_1^\pm
H^\mp$ open up] before it drops out because of the gradually closing phase
space for the decays $\tilde{g} \to q\bar{q} \chi^0_{3,4}, qq' \chi_2^\pm$. \s

In the non--universal case with $M_2=150$ GeV, the region with $\mu \lsim 200$
GeV is cut--away since the mass difference between the heavier and lighter 
$\chi$ states is too small for the decays into $H^\pm$ bosons to occur. 
For larger $\mu$, the situation becomes similar to the universal case and 
$\sigma \times {\rm BR}(H^\pm)$ values of the order of 0.3 pb can be
reached for not too large $H^\pm$ masses and favorable regions of parameter 
space, leading to a sample of a hundred thousand charged Higgs boson events 
for the luminosity expected at the LHC.

\subsubsection*{3.2.3 The case of top squarks}

The cross section for top squark production at the LHC can be rather large when
$m_{\tilde{q}} \ge m_{\tilde{g}} \ge m_{\tilde{t}_1}+m_t$; in this case all
squarks and gluinos will decay into $\tilde{t}_1$ final states. The stops will
decay into heavier $\chi$ states, with rates shown in Fig.~3, and the latter
will possibly decay into lighter $\chi$ particles and $H^\pm$ bosons as shown
in Figs.~5 and 6. The situation will then be almost exactly similar to gluino
decays discussed in the previous subsection. The reason is that gluino decays
into the heavy $\chi$ particle occurred mainly through the virtual exchange of
$\tilde{t}_1$ and the branching ratios were controlled by the $\tilde{t}_1b
\chi_2^\pm$ and $\tilde{t}_1 t \chi_{3,4}^0$ vertices. The only difference with
the previous discussion will  be due to the smaller value of $m_{\tilde{t}_1}$
[since now $\tilde{t}_1$ is not virtual anymore] which will suppress the phase
space for $\tilde{t}_1 \to b \chi_{2}^+, t \chi_{3,4}^0$ decays. \s

Note that an additional contribution to the cross section will come from
the direct production of stop quarks $gg/q\bar{q}\to \tilde{t}_1\tilde{t}_1^*$,
which for $m_{\tilde{t}_1} \sim 500$ GeV can reach the picobarn level. This 
may substantially increase the number of charged Higgs bosons in the final 
state through the cascade decay $\tilde{t}_1 \to b\chi_2^+ \to b \chi_{1,2}^0 
H^+$ for instance.  

\section*{4. Direct $H^+$ decays of squarks and gluinos}

\subsection*{4.1 Two--body decays of squarks into Higgs bosons} 

If the mass splitting between two squarks of the same generation is large
enough, as is generally the case of the $(\tilde{t},\tilde{b})$ iso--doublet,
the heavier squark can decay into a lighter one plus a gauge boson $V=W,Z$ or a
Higgs boson $\Phi=h,H,A,H^\pm$. The partial decay widths are given at the
tree--level by [the QCD corrections to these decay modes have also been
calculated and can be found in Ref.~\cite{C5}]:
\beq
\Gamma(\tilde{q}_i \to \tilde{q}_j' V) &=& \frac{\alpha}{4 M_V^2} 
m_{\tilde{q}_i} \, g_{\tilde{q}_i \tilde{q}_j' V}^2 \, \lambda^{3/2} 
( \mu_V^2, \mu_{\tilde{q}_j'}^2) \\
\Gamma(\tilde{q}_i \to \tilde{q}_j' \Phi) &=& \frac{\alpha}{4} 
m_{\tilde{q}_i} \, g_{\tilde{q}_i \tilde{q}_j' \Phi}^2 \, \lambda^{1/2} 
(\mu_{\Phi}^2, \mu_{\tilde{q}_j'}^2) 
\eeq
In these equations, the couplings of the Higgs bosons to squarks, 
$g_{\tilde{q}_i \tilde{q}_j' \Phi}$, read in the case of neutral Higgs bosons:
\beq
g_{\tilde{q}_1 \tilde{q}_2 h} &=& \frac{1}{4s_W M_W} \, \bigg[ 
M_Z^2 s_{2\theta_q} (2I_q^3-4 e_q s_W^2) \sin (\alpha+\beta) 
+2 m_q c_{2\theta_q} (A_q r_2^q + 2I_q^3 \, \mu \, r_1^q) \bigg] \non \\
g_{\tilde{q}_1 \tilde{q}_2 H}&=& \frac{1}{4s_W M_W} \, \bigg[ 
- M_Z^2 s_{2\theta_q} (2I_q^3-4 e_q s_W^2) \cos (\alpha+\beta) 
+2 m_q c_{2\theta_q} (A_q r_1^q - 2I_q^3 \, \mu \, r_2^q) \bigg] \non \\
g_{\tilde{q}_1 \tilde{q}_2 A} &=& - g_{\tilde{q}_2 \tilde{q}_1 A}= 
\frac{-m_q}{2s_W M_W} \, \bigg[ \mu +A_q (\tan \beta)^{-2I_q^3} \bigg] 
\eeq
with the coefficients $r^q_{1,2}$ as [$\alpha$ is a mixing angle in the 
CP--even Higgs sector of the MSSM, and at the tree--level, can be expressed 
only in terms of $M_A$ and $\tan \beta$]
\beq
r_{1}^t = \frac{ \sin \alpha}{\sin \beta} \ \  , \ \ 
r_{2}^t = \frac{ \cos \alpha}{\sin \beta} \ \ , \ \
r_{1}^b = \frac{ \cos \alpha}{\cos \beta} \ \ , \ \ 
r_{2}^b = - \frac{ \sin \alpha}{\cos \beta}\;.
\eeq
In the case of the charged Higgs boson, the couplings to squarks are given,
in terms of the squark mixing matrices $ R^{\tilde{q}}$, by
\beq
\label{couplingmatrix}
g_{\tilde{q}_i \tilde{q}_j' H^\pm}= \frac{1}{2s_W M_W} \, \sum_{k,l=1}^2 \  
\left( R^{\tilde{q}} \right)_{ik} \, C_{ \tilde{q}
\tilde{q}' H^\pm }^{kl} \, \left( R^{\tilde{q}'} \right)_{lj}^{\rm T}
\eeq
with the matrix $C_{\tilde{q} \tilde{q}' H^\pm }$ summarizing the couplings 
of the $H^\pm$ bosons to the squark current eigenstates and it is given by 
\beq
C_{\tilde{t} \tilde{b}H^\pm} &= & \sqrt{2} \, \left( \begin{array}{cc}
m_b^2 \tb + m_t^2/\tb - M_W^2 \, \sin 2\beta & m_b \,(A_b \tb  +\mu) \\
m_t \,(A_t/\tb  +\mu) & 2  \,m_t \,m_b/ \sin2\beta
             \end{array} \right)
\eeq
For the couplings of squarks to the $W$ and $Z$ gauge bosons, one has 
\beq
\label{Z0couplings}
g_{\tilde{q}_1 \tilde{q}_2 Z}
&=& g_{\tilde{q}_2 \tilde{q}_1 Z} =  \frac{2I_q^3 s_{2\theta_q}}{4s_W c_W} 
\non \\
g_{\tilde{q}_i \tilde{q}_j' W} & = & \frac{1}{\sqrt{2} s_W} \left( 
\begin{array}{cc}
     \ct{q} \ct{q'} & - \ct{q} \st{q'} \\ -\st{q} \ct{q'} & \st{q} \st{q'}
  \end{array} \right)  
\eeq

\begin{figure}[htbp]
\begin{center}
\vskip-5cm
\hskip-2cm\centerline{\epsfig{file=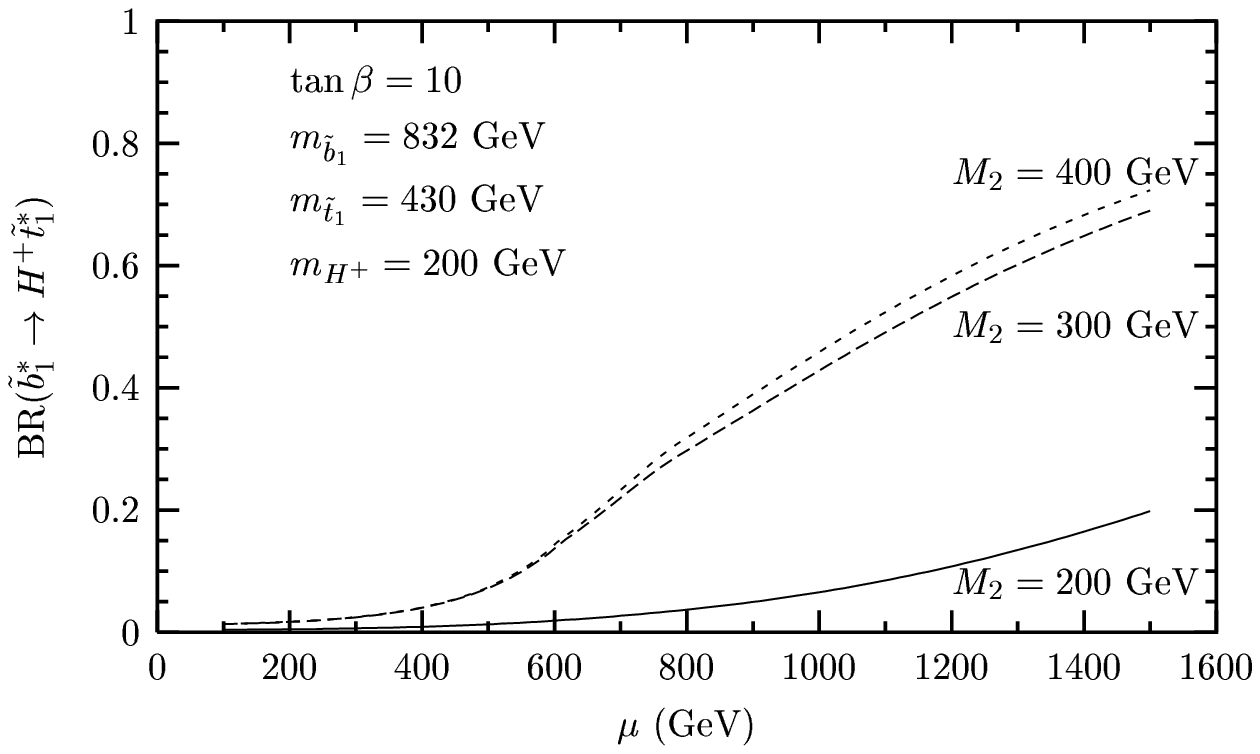,width=20cm}}
\vskip-20.5cm
\hskip-2cm\centerline{\epsfig{file=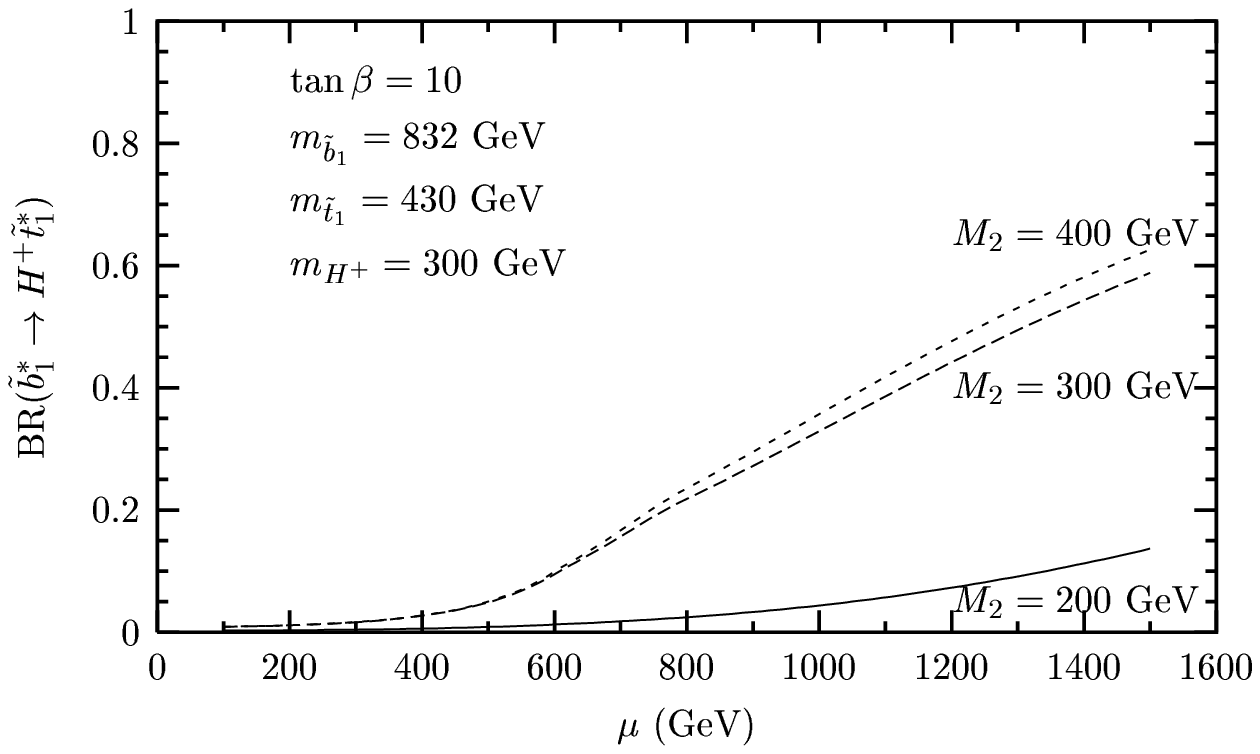,width=20cm}}
\vskip-17cm
\end{center}
\caption{The branching ratios for bottom squarks decaying into top squarks
and charged Higgs bosons as a function of $\mu$ for $\tb=10$ and $M_2=200,
300$ and 400 GeV. The charged Higgs boson mass is taken to be $M_{H^\pm}=
200$ and 300 GeV, in the upper and lower panels, respectively. The two squark 
masses are taken to be $m_{\tilde{b}_1}=832$ GeV and $m_{\tilde{t}_1}=430$ 
GeV.}
\vskip-1cm
\end{figure}

In Fig.~9, we display the branching ratio for the decays of a bottom squark
into the lightest top squark and a charged Higgs boson, $\tilde{b}_{1} \to
\tilde{t}_1 H^-, \tilde{b}_1^* \to \tilde{t}_1^* H^+$ as a function of the
parameter $\mu$ with three values of $M_2=200, \, 300$ and 400 GeV. We have
fixed $\tb=10$ and the sbottom, stop and charged Higgs boson masses to
$m_{\tilde{b}_1}=832$ GeV, $m_{\tilde{t}_1}=430$ GeV and $M_{H^\pm}=200 (300)$
GeV in the upper (lower) panel. The other competing decays of sbottoms are
decays into bottom quarks and neutralinos and top quarks and charginos as well
as bottom and gluino final states. We have assumed universal gaugino masses so
that the gluino mass is given by $m_{\tilde{g}} \sim 3M_2$; the strong
interaction decay channel $\tilde{b}_{1} \to b \tilde{g}$ is therefore only
open for $M_2=200$ GeV and is dominant in this case. \s

As can be seen, for $M_2 \ge 300$ GeV [i.e when there is no phase space for the
decay channel $\tilde{b}_{1} \to b \tilde{g}$ to occur],
BR($\tilde{b}_{1}\to\tilde{t}_1 H^-)$ can be substantial for large $\mu$
values, $\mu \gsim 700$ GeV, possibly exceeding the level of 50\%. The reason
for this feature, besides the fact that for $\mu \gsim 800$ GeV, the
$\tilde{b}_1$ decays into the heavier chargino and neutralinos are
kinematically closed, is that the sbottom--stop--$H^\pm$ coupling is strongly
enhanced and becomes larger than the sbottom--bottom--gaugino coupling which
controls the sbottom decays into the lighter chargino and neutralinos. For
smaller values of $M_2$, as pointed out earlier,  the decay $\tilde{b}_{1} \to
b \tilde{g}$ becomes accessible and would be the dominant decay channel. \s

Another possibility would be the decays of the heaviest top squark into a 
sbottom quark and charged Higgs bosons, $\tilde{t}_2 \to \tilde{b}_{1,2} H^+$: 
since the $\tilde{t}_2-\tilde{t}_1$ mass splitting can be large, there might
be enough phase space for this process to occur. However, the cross sections
for $\tilde{t}_2$ from direct production are not large, and if the gluinos
are heavier than squarks, they will decay into $\tilde{t}_2 t$ final states 
only less than $\sim 10 \%$ of the time. This decay channel is therefore less 
favored than the previous one. \s 

Note that for light top squarks, $m_{\tilde{t}_1} \lsim m_t +m_{\chi_1^0}$ and
$m_b+m_{\chi_1^\pm}$, and light charged Higgs bosons $M_{H^\pm} \lsim
m_{\tilde{t}_1}-m_{\chi_1^0}$, the three body decay $\tilde{t}_1 \to
bH^+\chi_1^0$ [which is mediated by virtual chargino, top or sbottom exchange]
is accessible and would compete with the other possible three--body decay mode
$\tilde{t}_1 \to bW^+\chi_1^0$ and the loop induced and flavor changing decay
$\tilde{t}_1 \to c \chi_1^0$.  In some areas of the parameter space, the stop
decay into charged Higgs bosons can be dominant; see the discussions in
Ref.~\cite{three-body}.

\subsection*{4.2 Three body decays of gluinos into charged Higgs bosons}

Finally, we discuss the direct decays of gluinos into top squarks, bottom 
quarks and charged Higgs bosons, mediated by virtual top quark or bottom 
squark exchanges; Fig.~10. 

\vspace*{-1.cm}
\begin{picture}(1000,200)(10,0)
\Text(140,130)[]{$\tilde{g}$}
\ArrowLine(110,120)(160,120)
\DashArrowLine(160,120)(190,150){4}
\Text(197,150)[]{$\tilde{t}_1$}
\ArrowLine(160,120)(180,90)
\Text(155,100)[]{$\bar{t}$}
\ArrowLine(180,90)(210,120)
\Text(215,120)[]{$\bar{b}$}
\DashArrowLine(200,65)(180,90){4}
\Text(210,65)[]{$H^-$}
\Text(300,130)[]{$\tilde{g}$}
\ArrowLine(260,120)(310,120)
\ArrowLine(310,120)(340,150)
\Text(347,150)[]{$\bar{b}$}
\DashArrowLine(310,120)(330,90){4}{}
\Text(305,100)[]{$\tilde{b}_i$}
\DashArrowLine(330,90)(360,120){4}
\Text(370,120)[]{$\tilde{t}_1$}
\DashArrowLine(350,65)(330,90){4}
\Text(360,65)[]{$H^-$}
\end{picture}
\vspace*{-2.5cm}

\begin{center}
\nn {Figure 10: The Feynman diagrams contributing to the three--body decay 
$\tilde{g} \to \tilde{t}_1 \bar{b} H^-$.} 
\end{center}
\vspace*{4mm} 
\setcounter{figure}{10}

The Dalitz density for this decay mode, taking into account all the masses of 
the final state particles, including the bottom quark, is given by: 
\beq
\frac{ \dx \Gamma} {\dx x_1 \dx x_2} (\tilde{g} \to H^- \bar{b} \tilde{t}_1)
&=& \frac{\alpha \alpha_s}{64 \pi} m_{\tilde{g}} \,  \bigg[ {\rm d}\Gamma_{t} + 
{\rm d}\Gamma_{\tilde{b}} + 2 {\rm d}\Gamma_{t\tilde{b}} \bigg]
\eeq
In terms of $x_1=2E_{H^\pm}/m_{\tilde{g}}, \, x_2=2E_{b}/m_{\tilde{g}}$ and the
reduced masses $\mu_X=m_X/m_{\tilde{g}}$, the squared $t$, $\tilde{b}$ 
contributions and the $t\tilde{b}$ interference are given by
\beq
{\rm d}\Gamma_{t} &=& 
\frac{1}{(x_1+x_2-1+ \mu_{\tilde{t}_1}-\mu_t)^2} \bigg\{
\mu_t x_2 \left[ y_t^2 c_\beta^2 (b^{\tilde t}_{1\tilde g})^2+
                 y_b^2 s_\beta^2 (a^{\tilde t}_{1\tilde g})^2 \right] \non \\
&& +\left[ y_t^2 c_\beta^2 (a^{\tilde t}_{1\tilde g})^2+
                 y_b^2 s_\beta^2 (b^{\tilde t}_{1\tilde g})^2 \right]\left[
x_1(1-\mu_b+\mu_{H^+}-\mu_{\tilde t_1}-x_1-x_2)+x_2(\mu_{H^+}-\mu_b)
\right] \bigg\} \non \\
{\rm d}\Gamma_{\tilde{b}} &=& \sum_{i,j=1}^2
\frac{g_{\tilde t_1\tilde b_i H^+}g_{\tilde t_1\tilde b_j H^+}
[a^{\tilde b}_{i \tilde g} a^{\tilde b}_{j \tilde g}
+b^{\tilde b}_{i \tilde g} b^{\tilde b}_{j \tilde g}]x_2}
{(1+\mu_b-x_2-\mu_{\tilde b_i})(1+\mu_b -x_2 - \mu_{\tilde b_j})} \non \\
{\rm d}\Gamma_{t\tilde{b}_1} &=& \sum_{i=1}^2 \frac{g_{ \tilde{t}_1 
\tilde{b}_H^+} } {(1+\mu_b-x_2-\mu_{\tilde b_i})
(x_1+x_2-1+ \mu_{\tilde t_1} -\mu_t)} \bigg\{
-\sqrt{\mu_t} x_2 \left[ y_t c_\beta b^{\tilde b}_{i \tilde g} 
b^{\tilde t}_{1 \tilde g} +y_b s_\beta a^{\tilde b}_{i \tilde g}
a^{\tilde t}_{1 \tilde g} \right] \non \\
&&+ [ y_t c_\beta b^{\tilde b}_{i \tilde g} a^{\tilde t}_{1 \tilde g}
+ y_b s_\beta a^{\tilde b}_{i \tilde g}b^{\tilde t}_{1 \tilde g} ]
(x_1+x_2-1-\mu_b-\mu_{H^+}+\mu_{\tilde t_1}) \bigg\}
\eeq
The various couplings have been given in the previous sections and the Yukawa
couplings of top and bottom quarks are given by: $y_t=m_t/(\sqrt{2}s_W M_W
s_\beta)$ and $y_b=m_b/(\sqrt{2}s_W M_W c_\beta)$.  \s

To obtain the partial decay width, one has to integrate over $x_1$ and $x_2$ 
with boundary conditions: 
\beq
2 \sqrt{\mu_{H^+}} \ \leq \ x_1 \ \leq 1+ [\mu_{H^+} -(\sqrt{\mu_b}+ \sqrt{
\mu_{\tilde{t}_1}} )^2] 
\eeq
\beq
s_{\rm min} \ \leq x_2 \ \leq s_{\rm max} 
\eeq
with 
\beq
s_{\rm min} &=& \frac{1}{2} \frac{(x_1-2)(x_1 -1 -\mu_b + \mu_{\tilde{t}_1} -
\mu_{H^+} ) - \sqrt{\Delta} }{1-x_1+ \mu_{H^+} } \non \\
s_{\rm max} &=& \frac{1}{2} \frac{(x_1-2)(x_1 -1 -\mu_b + \mu_{\tilde{t}_1} -
\mu_{H^+} ) +\sqrt{\Delta} }{1-x_1+ \mu_{H^+} } 
\eeq
and
\beq
\Delta =   (\mu_{H^+} -x_1^2) \left[ \frac{1}{4} \mu_b \mu_{\tilde{t}_1} - 
(x_1 -1 + \mu_b +\mu_{\tilde{t}_1} - \mu_{H^+} )^2 \right]
\eeq
The branching fraction for the three--body decay, BR($\tilde{g} \to \tilde{t}_1
\bar{b}H^- + \tilde{t}_1^* bH^+)$ is illustrated in Fig.~11 as a function of
$\mu$ for $\tan \beta=10$. We have chosen squark masses of $m_{\tilde{q}}
=m_{\tilde{b}_i}=1$ TeV, a gluino mass slightly lower, $m_{\tilde{g}}=900$ GeV,
and the lighter stop mass to be $m_{\tilde{t}_1}=433$ GeV; for the charged
Higgs boson mass we take three values: $M_{H^\pm}=190, 230$ and 310 GeV. 
Thus in this scenario, all squarks [including bottom squarks] will decay into 
almost massless quarks and gluinos and the latter will dominantly decay into
the lighter top squarks and top quarks. The three--body decays $\tilde{g} \to 
\tilde{t}_1 \bar{b}H^-$ and $\tilde{g} \to \tilde{t}_1^* bH^+$ have therefore
to compete with a strong interaction two--body decay, which has a large phase
space in this case. This is the reason why the branching ratio hardly
exceeds the one percent level, which occurs for large $\mu$ values 
when the $\tilde{t}\tilde{b}H^\pm$ couplings are enhanced. Note that the 
smallness of the branching ratio is also due to the smallness of the $tbH^+$
coupling for the chosen value of $\tb$; for larger or smaller values
of $\tb$, the branching ratio can be significantly larger [e.g. an order of
magnitude larger than in Fig.~11 for $\tb=30$]. \s

In spite of the small branching ratio, the number of $H^\pm$ final states
can be rather large at the LHC in the kinematical configuration shown in 
Fig.~11. Indeed, for the chosen squark and gluino masses, the cross section
for gluinos coming from direct production and from squark decays is at the
level of $\sim 5$ picobarn. With branching fractions of the order of a few 
percent, this means that the cross section for $H^\pm$ production via gluino 
decays can reach the level of a fraction of a  picobarn leading, in the 
favorable case, to a sample of a few ten thousand $H^+$ events with an 
integrated luminosity of $\int {\cal L}=300$ fb$^{-1}$. \s

\begin{figure}[htbp] 
\begin{center} 
\vskip-5cm
\hskip-2cm\centerline{\epsfig{file=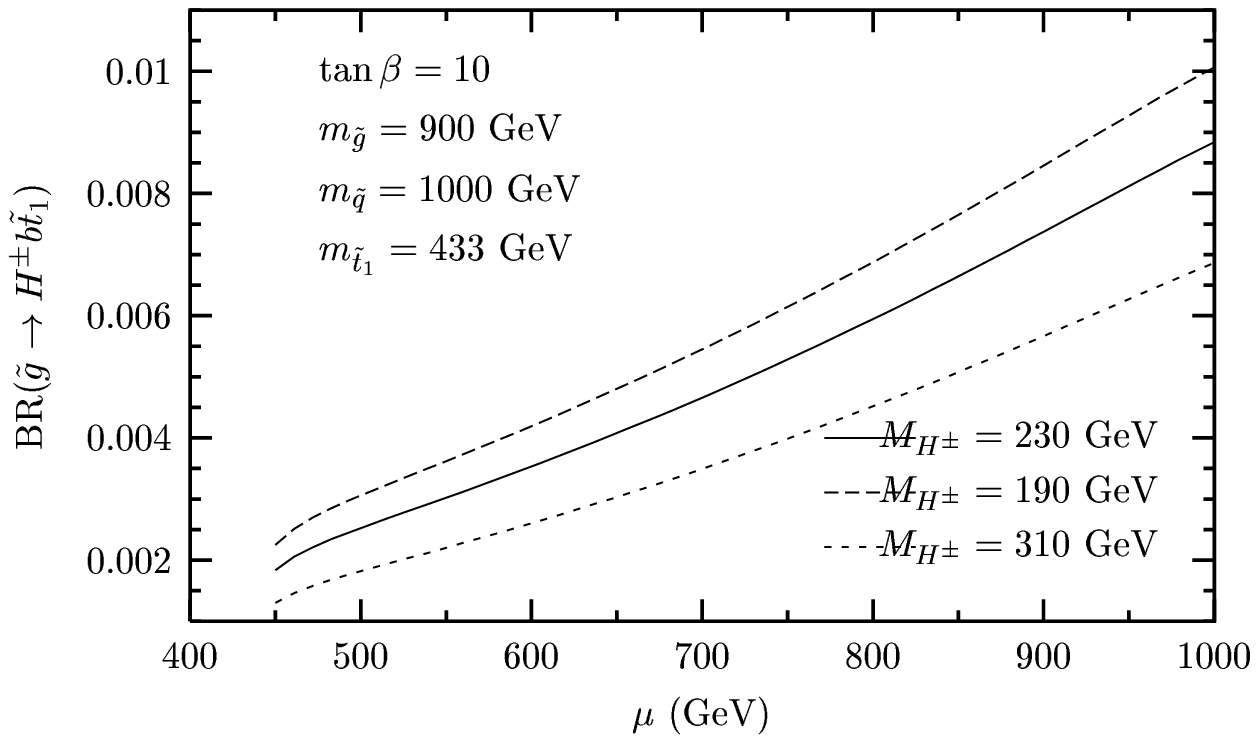,width=20cm}} 
\vskip-17cm
\end{center} 
\caption{The branching ratios for direct decays of gluinos into
bottom quarks, top squarks and charged Higgs bosons as a function of $\mu$ for
$\tb=10$ and $M_{H^\pm}=190,230$ and 310 GeV. The squark masses are taken to be
$m_{\tilde{q}}=1$ TeV and $m_{\tilde{t}_1}=433$ GeV, while the gluino mass is
$m_{\tilde{g}}=900$ GeV.} 
\vspace*{.1cm} 
\end{figure} 

Finally, let us note that for charged Higgs bosons lighter than top quarks, the
decay $t \to bH^+$ will occur at the two--body level and the branching fraction
can be rather large [of the order of 10 percent for $\tb=10$]. In the
kinematical configuration discussed often in this analysis, i.e. squarks
heavier than gluinos which are heavier than $m_{\tilde{t}_1}+m_t$, all strongly
interacting particles will cascade into stop and top quarks, with the latter
decaying into charged Higgs bosons and bottom quarks. The number of $H^+$ boson
can be therefore substantial; see also the discussion given in \cite{Baer}.
These processes will complement [and possibly, also compete with] searches of
$H^\pm$ bosons from top quark decays with these particles produced directly in
$pp$ collisions.  
 
\section*{5. Conclusions} 

We have analyzed the production of the charged Higgs particles of the
unconstrained Minimal Supersymmetric Standard Model from cascade as well as
direct decays of squarks and gluinos, which have large production cross
sections at the LHC due to their strong couplings.  \s 

Squarks and gluinos can decay into the heavier chargino and neutralinos, which
subsequently decay into the lighter chargino and neutralino states and charged
Higgs bosons. We have shown that the branching fractions for these decay modes
can be rather large in some regions of the MSSM parameter space, and that the
cross sections times the branching ratios for $H^\pm$ production can be
significantly large. This would allow  the possibility to produce large samples
of these states at the LHC, in particular in situations [$H^\pm$ masses larger
than the top quark mass or intermediate values of $\tb$, for instance] where
their production cross sections in the direct processes [associated production
with bottom quarks or with $tb$ final states] are rather small. \s

We have also discussed the possibility of producing charged Higgs bosons either
through direct decays of the scalar partners of heavy quarks, as is the case in
the process $\tilde{b}_{1,2} \to H^-\tilde{t}_1$, or in direct three--body
decays of gluinos, $\tilde{g} \to \tilde{t}_1 \bar{b}H^-$ and $\tilde{t}_1^*
bH^+$.  In some favorable regions of the parameter space, the branching ratios 
of these decay channels, in particular in the former case, can be large enough 
to allow for the detection of the charged Higgs bosons. For gluino decays, the
situation becomes even more favorable for light $H^\pm$ bosons when the
two-body decays $t \to bH^+$ are kinematically possible, leading to a surplus
of events compared to the case $pp \to \bar{t}t \to H^\pm +X$. \s

Thus, there is an additional source of charged Higgs particles at the LHC with
interesting signals\footnote{We thank Filip Moortgat from the CMS collaboration
for a discussion on this point.} in most cases, since the final states involve
missing energy, multi--leptons [from the decay cascades] and heavy flavors [$b$
and $t$ quarks]. These signals would help in detecting these particles in the
difficult environment of the LHC. In our parton level analysis of these
process, we did not attempt to scan the entire MSSM parameter space, discuss
the final state topologies and the possible backgrounds, etc., although for the
cross sections and branching ratios, we tried to be rather conservative [for
instance by not including $K$--factors, etc.]. A detailed analysis taking into
account all backgrounds, selection and detection efficiencies in a realistic
way, which is beyond the scope of the present paper, is required to assess in
which part of the MSSM parameter space these final states can be isolated
experimentally.  

\bigskip

\nn {\bf Acknowledgments}: \s

\nn We thank  Daniel Denegri, Filip Moortgat and some members of the {\it
Beyond the Standard Model} group at the Workshop {\it Physics at TeV colliders}
in Les Houches (June 2001), for discussions.  M. Guchait thanks the CNRS for
financial support and the LPMT for the hospitality extended to him. A.~Datta
and Y. Mambrini are supported by a MNERT fellowship and Y.M. thanks the members
of the SPN for the facilities accorded to him. This work is supported in part
by the GDR--Supersym\'etrie and by the European Union under contract
HPRN-CT-2000-00149.

\end{document}